\documentclass[smallextended]{svjour3}     
\smartqed  
\usepackage{graphicx}
\usepackage{latexsym,amsmath,amssymb,bm,longtable}
\usepackage{subfigure}
\journalname{Journal of Statistical Physics}

\begin{document}

\title{A simple kinetic model for the phase transition of the van der Waals fluid }

\author{Shigeru Takata         \and
        Takashi Noguchi 
}

\institute{S. Takata \at
              Department of Aeronautics and Astronautics,
              Kyoto University, Kyoto 615-8540, Japan \\
              \email{takata.shigeru.4a@kyoto-u.ac.jp}           
            \and
           T. Noguchi \at
              Department of Aeronautics and Astronautics,
              Kyoto University, Kyoto 615-8540, Japan 
}

\date{\today}
\maketitle

\begin{abstract}
A simple kinetic model, which is presumably minimum, for the phase
transition of the van der Waals fluid is presented. In the model,
intermolecular collisions for a dense gas has not been treated faithfully.
Instead, the expected interactions as the non-ideal gas effect are
confined in a self-consistent force term. Collision term plays just
a role of thermal bath. Accordingly, it conserves neither momentum
nor energy, even globally. It is demonstrated that (i) by a natural
separation of the mean-field self-consistent potential, the potential
for the non-ideal gas effect is determined from the equation of state
for the van der Waals fluid, with the aid of the balance equation
of momentum, (ii) a functional which monotonically decreases in time 
is identified by the H theorem and 
is found to have a close relation to the Helmholtz free energy in
thermodynamics, and (iii) the Cahn\textendash Hilliard type equation
is obtained in the continuum limit of the present kinetic model. Numerical
simulations based on the Cahn\textendash Hilliard type equation are
also performed.
\end{abstract}

\vspace{2pc}

\makeatletter

\section{Introduction}

It is well-known that gas behavior in both equilibrium and non-equilibrium
states is well described by the kinetic theory of gases or the Boltzmann
equation. In the Boltzmann equation, short-range molecular interactions
are treated as instantaneous binary collision events between sizeless
particles, and accordingly it is applied to ideal (or perfect) gases.
The first attempt to deal with the non-ideal gas effect in the framework
of kinetic theory goes back to the dates of Enskog \cite{CC95,HCB64}.
In his celebrated equation, the displacement effect in collision events
is considered, leading to instantaneous transfer of momentum
and energy in a molecular-size distance. Some authors make use of
the Vlasov\textendash Enskog equation \cite{G71,FGL05,KOW12} 
for the study of liquid-vapor phase transition. 
In this equation the collision dynamics
of the Enskog model is retained and long-range interactions are dealt
with by a collective mean field, like in the Vlasov (or Vlasov\textendash Poisson) equation for plasma.
Some recent research trends in the connection to the kinetic theory
for both gas and liquid phases can be found, e.g., in \cite{FB17}.

The above mentioned approaches are quite reasonable. For our primary
concern, however, it contains too much details of the molecular scale information. 
We are not necessarily interested in full details in that scale
but rather interested in the dynamics of phase transition and a simple kinetic theory description for it.
We require to such a theory a capability of describing gas flows 
far out of equilibrium
near the liquid interface and hopefully simplifying the descriptions in the recent literature.
In this sense, our aim falls into the category of  
the original kinetic theory extension like \cite{FGL05,KOW12}.
It is in its philosophy different from many proposals 
in the framework of the lattice Boltzmann method, e.g., \cite{SOOY96,GLS07},
because they are naturally limited to the continuum regime and thus to weakly nonequilibrium setting. 

In the present paper, we introduce the simplest version of our model.
This is the first step of our approach toward the construction of
kinetic model equipped with the above mentioned capability that we want.
In this version, 
full details of intermolecular collisions for the non-ideal gas are not considered; 
the collision term plays a role
just as a thermal bath and conserves neither momentum nor energy, even globally. 
The expected interactions that induce non-ideal gas effects are simply collected
into a self-consistent force field. 
We stress that, even with this simplest version,
the essential features of phase transition dynamics can be recovered, as will be shown both theoretically
and numerically in sections~\ref{subsec:Another-viewpoint-based} and
\ref{subsec:Numerical-simulation-of}.
We here mainly show that (i) by a natural separation
of the mean-field self-consistent potential, the potential for the
non-ideal gas effect is determined from the equation of state for
the van der Waals fluid, with the aid of the balance equation of momentum,
(ii) a functional which monotonically decreases in time 
is identified from the H theorem and is found to have a close relation to the Helmholtz free energy in
thermodynamics, and (iii) the Cahn\textendash Hilliard type equation is obtained in
the continuum limit of the present kinetic model. 
The last item (iii) is a natural consequence of  the dissipative nature of the collision term.
Some results of numerical simulations based on the obtained Cahn\textendash Hilliard
type equation will be presented as well.

\section{Thermal bath and self-consistent mean field \label{sec:Heat-bath-and}}

We are going to consider the following kinetic equation for a system
composed of innumerable molecules in a periodic spatial domain $D$:%
\begin{subequations}
\begin{align}
 \frac{\partial f}{\partial t} +
&\xi_{i}\frac{\partial f}{\partial X_{i}}
+F_{i}\frac{\partial f}{\partial\xi_{i}}
=C_{*}[f],\label{eq:kinetic}\displaybreak[0]\\
&C_{*}[f]=A(\rho)(\rho M_{*}-f),\quad A(\rho)>0,\label{eq:col}\displaybreak[0]\\
&\rho[f]=\int fd\bm{\xi},\quad F_{i}=-\frac{\partial\phi}{\partial X_{i}},\quad\phi=\Phi_{S}(\rho)+\Phi_{L}[\rho], \label{eq:mom}\displaybreak[0]\\
&M_{*}=\frac{1}{(2\pi RT_{*})^{3/2}}\exp(-\frac{\xi^{2}}{2RT_{*}}), \label{eq:max}
\end{align}%
\end{subequations}%
where $t$ is a time, $\bm{X}$ a position, $\bm{\xi}$
a molecular velocity, $\xi=|\bm{\xi}|$, $f(t,\bm{X},\bm{\xi})$ a
velocity distribution function (VDF), $m\bm{F}$ a force acting on
a single molecule, with $m$ being its mass, and $\phi$ its corresponding
potential. $C_{*}[f]$ is a so-called collision term and plays a role
of a thermal bath and drives the system toward the thermal equilibrium
at temperature $T_{*}$. $A$ is assumed to be a positive function
of the local density $\rho$ and $R=k_{B}/m$ with $k_{B}$ being
the Boltzmann constant. We distinct two types of brackets $(\cdot)$
and $[\cdot]$ in the above: the former represents the argument of
a function, while the latter represents that of a functional or an
operator. The range of intergration $\mathbb{R}^3$ with respect to 
$\bm\xi$ (and its dimensionless counterpart $\bm\zeta$) will be omitted in the present paper, following the convention in nonmathematical literature. Einstein's notation on repeated indexes will be used throughout the present paper. 
Some explanation of the splitting of $\phi$ into $\Phi_{S}$
and $\Phi_{L}$ would be in order. 

The self-consistent force potential $\phi$ is split into attractive
and repulsive parts. The attractive part, $\Phi_{A}$, is of long-range,
while the repulsive part, $\Phi_{R}$, is of short-range and is a
function of the local density $\rho$. By the latter and a part of
the former, we intend to reproduce a non-ideal gas feature under the
isothermal approximation, which is represented by the potential $\Phi_{S}$.
Excluding effect by the repulsive force is usually included in the
collision term with detailed collision dynamics, like in the Enskog
equation\cite{CC95,HCB64}. Hence, the simplification by combining
the mean-field repulsive potential and the simplified role of the
collision term is the main difference from the existing model \cite{G71,FGL05}.

The attractive mean field is expressed by
\begin{align}
 m\Phi_{A}(t,\bm{X}) 
&=\int_{\mathbb{R}^3}\Psi(|\bm{r}|)\{\rho(t,\bm{X}+\bm{r})-\rho(t,\bm{X})\}d\bm{r}
 +\int_{\mathbb{R}^3}\Psi(|\bm{r}|)d\bm{r}\rho(t,\bm{X})\nonumber \\
& \equiv m\Phi_{L}[\rho]+\int_{\mathbb{R}^3}\Psi(|\bm{r}|)d\bm{r}\ \rho(t,\bm{X}),
\label{eq:longrange}
\end{align}
where $m\Psi$ is the attractive intermolecular potential and is assumed
to be isotropic. Here, $\Phi_{L}$ may be considered as a contribution
from the long tail to the total attractive potential. The subtracted
part $\int_{\mathbb{R}^3}\Psi(|\bm{r}|)d\bm{r}\ \rho(t,\bm{X})$ will be combined
with the repulsive part to form the residue $m\Phi_{S}$ in the total
self-consistent potential $m\phi$:
\begin{equation}
m\Phi_{S}=m\Phi_{R}+\{\int_{\mathbb{R}^3}\Psi(|\bm{r}|)d\bm{r}\}\rho(t,\bm{X}),
\end{equation}
the functional form of which will be determined later from the van
der Waals equation of state in section~\ref{sec:Balance-equations-and}.
Since $\Phi_S$ is of short range, we are motivated to treat this as a local
 (or internal) variable, the stress tensor. This is the key idea behind our phenomenological determination of $\Phi_S$ from the equation of state 
(see section~\ref{sec:Balance-equations-and} for details). 
With the potential information thus determined, 
the above system (\ref{eq:kinetic})--(\ref{eq:max}) is closed.

When $\Psi$ decays fast in the system size as usually expected, the
variation of $\rho$ is moderate in that scale and the Taylor expansion
is allowed to yield
\begin{align}
  \Phi_{L}[\rho](t,\bm{X})
&=\frac{1}{m}\int_{\mathbb{R}^3}\Psi(|\bm{r}|)\{\rho(t,\bm{X}+\bm{r})-\rho(t,\bm{X})\}d\bm{r}\nonumber \\
&=\frac{1}{m}\int_{\mathbb{R}^3}\Psi(|\bm{r}|)\{r_i\frac{\partial}{\partial X_i}\rho(t,\bm{X})+\frac{1}{2}r_{i}r_{j}\frac{\partial^2}{\partial X_i\partial X_j}\rho(t,\bm{X})+\cdots\}d\bm{r}\nonumber \\
&\simeq\frac{1}{6m}\int_{\mathbb{R}^3}\Psi(|\bm{r}|)r^{2}d\bm{r}\frac{\partial^2}{\partial X_i^2}\rho(t,\bm{X})\equiv-\kappa\frac{\partial^2}{\partial X_i^2}\rho(t,\bm{X}).
\end{align}
Here $\kappa>0$, since $\Psi$ is attractive. The reduction from
the second to the last line is a consequence of the isotropic assumption
on $\Psi$.

\section{Balance equations and short range potential\label{sec:Balance-equations-and}}

Let us use the notation $\langle\cdot\rangle=\int\,\cdot\,d\bm{\xi}$
and define the flow velocity $v_{i}$ by $\rho v_{i}=\langle\xi_{i}f\rangle$.
By taking the $1$ and $\xi_{j}$-moments of (\ref{eq:kinetic}),
the balance equations of mass and momentum are obtained:%
\begin{subequations}\label{balance1} 
\begin{align}
 &\frac{\partial}{\partial t}\rho+\frac{\partial}{\partial X_i}(\rho v_{i})=0,\\
 &\frac{\partial}{\partial t}(\rho v_{j})+\frac{\partial}{\partial X_i}\langle\xi_{i}\xi_{j}f\rangle+\rho\frac{\partial \phi}{\partial X_j}=-A(\rho)\rho v_{j}.\label{bl1_2}
\end{align}
\end{subequations}%
Although we do not show it here, the balance equation
of energy is obtained as well by taking $\xi^{2}$-moment of (\ref{eq:kinetic}).
With the notation $c_{i}=\xi_{i}-v_{i}$ and the following reduction of the third term of (\ref{bl1_2})
\begin{align}
 \rho\frac{\partial}{\partial X_j}\phi
&
=\rho\frac{\partial}{\partial X_j}\{\Phi_{S}(\rho)+\Phi_{L}\}
=\rho\Phi_{S}^{\prime}\frac{\partial \rho}{\partial X_j}
+\rho\frac{\partial \Phi_{L}}{\partial X_j} \nonumber \\
&
=\frac{\partial}{\partial X_j}(\int\rho\Phi_{S}^{\prime}d\rho)
+\rho\frac{\partial \Phi_{L}}{\partial X_j},
\end{align}
where $\Phi_{S}^{\prime}$ denotes the derivative of $\Phi_{S}$,
the above balance equations are recast as%
\begin{subequations}\label{balance}
\begin{align}
&\frac{\partial}{\partial t}  \rho+\frac{\partial}{\partial X_i}(\rho v_{i})=0,\label{eq:mass}\\
&\frac{\partial}{\partial t}  (\rho v_{j})+\frac{\partial}{\partial X_i}(\rho v_{i}v_{j}+\langle c_{i}c_{j}f\rangle+\int\rho\Phi_{S}^{\prime}d\rho\delta_{ij})+\rho\frac{\partial \Phi_{L}}{\partial X_j}=-A(\rho)\rho v_{j}.\label{eq:momentum}
\end{align}
\end{subequations}
Here and in what follows, unless otherwise stated, 
the integrals with respect to $\rho$ (and its dimensionless counterparts 
$\tilde{\rho}$ and $\chi$ that will appear later) are indefinite integrals. 

Now, let us assume the van der Waals fluid. Then, the equation of
state is given by \cite{RW02}
\begin{equation}
p=\frac{\rho RT}{1-b\rho}-\rho^{2}a,\label{eq:vdw}
\end{equation}
where $a$ and $b$ are positive constants. In the meantime, the observation
of the balance equation of momentum motivates us to define the stress
tensor $p_{ij}$ and pressure $p$ as%
\begin{subequations}\label{stresspressure}
\begin{align}
& p_{ij} =\langle c_{i}c_{j}f\rangle+\int\rho\Phi_{S}^{\prime}d\rho\delta_{ij},\\
& p =\frac13 \langle\bm{c}^{2}f\rangle+\int\rho\Phi_{S}^{\prime}d\rho=\rho RT+\int\rho\Phi_{S}^{\prime}d\rho,
\end{align}
where the following usual definition of temperature $T$ has been introduced
\begin{equation}
T=\frac{1}{3\rho R}\langle\bm{c}^{2}f\rangle.
\end{equation}
\end{subequations}%
With these in mind, we can identify the functional
form of $\Phi_{S}$, under the isothermal approximation $T=T_{*}$,
by the relation
\begin{equation}
\rho RT_{*}+\int\rho\Phi_{S}^{\prime}d\rho\equiv p=\frac{\rho RT_{*}}{1-b\rho}-\rho^{2}a=\rho RT_{*}+\frac{b\rho^{2}RT_{*}}{1-b\rho}-\rho^{2}a,
\end{equation}
namely 
\begin{equation}
\int\rho\Phi_{S}^{\prime}d\rho=\rho RT_{*}(\frac{1}{1-b\rho}-1)-\rho^{2}a=\frac{b\rho^{2}RT_{*}}{1-b\rho}-\rho^{2}a.
\end{equation}
Straightforward calculations lead to the following expressions:%
\begin{subequations}
\begin{align}
 \rho\Phi_{S}^{\prime} & =\frac{b\rho RT_{*}}{1-b\rho}+\frac{b\rho RT_{*}}{(1-b\rho)^{2}}-2\rho a,\\
\Phi_{S}^{\prime} & =\frac{bRT_{*}}{1-b\rho}+\frac{bRT_{*}}{(1-b\rho)^{2}}-2a,\\
 \Phi_{S} & = -RT_{*}\ln(1-b\rho)+(\frac{RT_{*}}{1-b\rho}-RT_{*})-2a\rho\nonumber \\
& = -RT_{*}\ln(1-b\rho)+\frac{b\rho RT_{*}}{1-b\rho}-2a\rho,\\
\int\Phi_{S}d\rho & = RT_{*}\frac{(1-b\rho)\ln(1-b\rho)}{b}-\frac{RT_{*}}{b}\ln(1-b\rho)-a\rho^{2}\nonumber \\
&  =  -\rho RT_{*}\ln(1-b\rho)-a\rho^{2}.
\end{align}
\end{subequations}%
In the equation for $\Phi_{S}$, the integration
constant has been chosen so that $\Phi_{S}$ vanishes in the low density
limit ($\rho\to0$). 

In the meantime, a thermodynamically consistent definition of the
specific internal energy $e$ is given by 
\begin{equation}
e=\int\rho^{-2}(p-T\frac{\partial p}{\partial T})d\rho+\frac{3}{2}RT,
\end{equation}
which leads to the following definition within the present isothermal
approximation:
\begin{align}
e&=\frac{1}{2\rho}\langle\bm{c}^{2}f\rangle
       +\int\rho^{-2}(1-T_{*}\frac{\partial}{\partial T_{*}})(\int\rho\Phi_{S}^{\prime}d\rho)d\rho
          \nonumber \\
&=\frac{1}{2\rho}\langle\bm{c}^{2}f\rangle-\rho^{-1}(1-T_{*}\frac{\partial}{\partial T_{*}})(\int\rho\Phi_{S}^{\prime}d\rho)+\int\rho^{-1}(1-T_{*}\frac{\partial}{\partial T_{*}})(\rho\Phi_{S}^{\prime})d\rho \nonumber \\
&=\frac{1}{2\rho}\langle\bm{c}{}^{2}f\rangle
        -(1-T_{*}\frac{\partial}{\partial T_{*}})(\Phi_{S}-\rho^{-1}\int\Phi_{S}d\rho )
        +(1-T_{*}\frac{\partial}{\partial T_{*}})\Phi_{S} \nonumber \\
&=\frac{1}{2\rho}\langle\bm{c}{}^{2}f\rangle+\rho^{-1}(1-T_{*}\frac{\partial}{\partial T_{*}})\int\Phi_{S}d\rho=\frac{3}{2}RT-a\rho.
\end{align}
In a similar way, a thermodynamically consistent definition of the
specific entropy $s$ leads to the following definition of $s$ within
the present isothermal approximation:%
\footnote{To reach this form, we have taken into account two thermodynamical relations 
$\partial s/\partial\rho =-\rho^{-2}\partial p/\partial T$ and $\partial s/\partial T=T^{-1}\partial e/\partial T$, where the pair of $\rho$ and $T$ are chosen as independent variables.
Within the isothermal approximation, the former is integrated in $\rho$ to yield 
$s=s_0(T)-R\ln\rho -\rho^{-1}\frac{\partial}{\partial T_*}\int \Phi_S d\rho$.
Then, the second thermodynamic relation determines $s_0$ as $s_0(T)=(3/2)R\ln T+\mathrm{const}$.
Note that the set of the first two terms of $s$ is identical to the specific entropy for monatomic ideal gases. }%
\begin{align}
s & \equiv  \frac{3}{2}R\ln T-R\ln\rho-\frac{1}{\rho}\int\frac{\partial\Phi_{S}}{\partial T_{*}}d\rho+\mathrm{const.} \nonumber \\
 & = \frac{3}{2}R\ln\frac{T}{T_{*}}-R\ln\frac{\rho}{\rho_{0}}-\frac{1}{\rho}\int\frac{\partial\Phi_{S}}{\partial T_{*}}d\rho \nonumber \\
 & = R\ln(T/T_{*})^{3/2}-R\ln(\rho/\rho_{0})+\frac{1}{T_{*}}(e-\frac{3}{2}RT-\frac{1}{\rho}\int\Phi_{S}d\rho) \nonumber \\
 & = R\ln(T/T_{*})^{3/2}-R\ln(\rho/\rho_{0})+R\ln(1-b\rho),\label{sdef}
\end{align}
where the constant on the first line is determined so that $s$ for
the ideal gas vanishes when its density and temperature are respectively
$\rho_{0}$ and $T_{*}$. Combining above two, we have a relation
that
\begin{align}
\frac{1}{2}\langle\bm{c}{}^{2}f\rangle+\int\Phi_{S}d\rho 
&=  \rho e+T_{*}\frac{\partial}{\partial T_{*}}\int\Phi_{S}d\rho \nonumber \\
&= \rho e+T_{*}\{-\rho s+\rho R\ln(T/T_{*})^{3/2}-\rho R\ln(\rho/\rho_{0})\} \nonumber \\
&= \rho(e-T_{*}s)+\rho RT_{*}\{\ln(T/T_{*})^{3/2}-\ln(\rho/\rho_{0})\} \nonumber \\
&= \rho\mathcal{A}+\rho RT_{*}\{\ln(T/T_{*})^{3/2}-\ln(\rho/\rho_{0})\},
\label{eq:Helmholtz}
\end{align}
where $\rho_0$ is a reference density 
and $\mathcal{A}(\equiv e-T_{*}s)$ is identified, within the isothermal
approximation, as the specific Helmholtz free energy. The above relation
is useful to have a physical interpretation of a functional which monotonically decreases in time in section~\ref{sec:H-theorem-and}.

\begin{remark}
Since we have retained the effect of long-range interaction as it is,
the long-range part is not necessarily local.
Accordingly, we have included only the short-range part into the definition of pressure and stress tensor.
If one assumes $\Phi_L=-\kappa(\partial^2\rho/\partial X_i^2)$ from the beginning,
the long-range part ought to be local as well and can be included into the pressure and stress tensor.
In the case, the third term on the right-hand side of \eqref{eq:defM} that appears later,
namely the interface energy, may be interpreted as the effect of additional stress term
which is appreciable only in a sharp change region, like the interface.
This type of interpretation corresponds to a phenomenological fluiddynamic approach that introduces an additional stress at the interface. 
Here we do not take this interpretation, since we treat the long-range interaction which is not necessarily local.
\end{remark}

\section{H theorem and Helmholtz free energy\label{sec:H-theorem-and}}

The collision operator $C_{*}$ plays a role of the thermal bath and
has a following property: 
\begin{align}
   \langle(1+\ln\frac{f}{\rho_{0}M_{*}})\,C_{*}[f]\rangle 
& =\langle\{1+\ln(\frac{\rho}{\rho_{0}})+\ln(\frac{f}{\rho M_{*}})\}\,A(\rho)(\rho M_{*}-f)\rangle\nonumber \\
 & =A(\rho)\rho\langle M_{*}(1-\frac{f}{\rho M_{*}})\ln\frac{f}{\rho M_{*}}\rangle\le0,
\end{align}
where the equality holds only when $f=\rho M_{*}$. The same operation
as above on the left-hand side of (\ref{eq:kinetic}) eventually leads to
\begin{align}
 \langle(1+&\ln\frac{f}{\rho_{0}M_{*}})(\frac{\partial f}{\partial t}+\xi_{i}\frac{\partial f}{\partial X_{i}}+F_{i}\frac{\partial f}{\partial\xi_{i}})\rangle \nonumber \\
= & \frac{\partial}{\partial t}\{\langle f\ln\frac{f}{c_{0}}\rangle+\rho\ln(\frac{T^{3/2}}{T_{*}^{3/2}}\frac{\rho_{0}}{\rho})+\frac{\rho}{RT_{*}}(\mathcal{A}+\frac{1}{2}v^{2})\}\nonumber \\
& +\frac{\partial}{\partial X_{i}}\Big\{\langle\xi_{i}f\ln\frac{f}{c_{0}}\rangle+\rho v_{i}\ln(\frac{T^{3/2}}{T_{*}^{3/2}}\frac{\rho_{0}}{\rho})+\frac{1}{RT_{*}}\{\rho(\mathcal{A}+\frac{1}{2}v^{2})v_{i}\nonumber \\
& \qquad+\frac{1}{2}\langle c_{i}\bm{c}^{2}f\rangle+p_{ij}v_{j}\}\Big\}+\frac{\rho v_{i}}{RT_{*}}\frac{\partial\Phi_{L}}{\partial X_i},
  \label{eq:H_left}
\end{align}
where $v=|\bm{v}|$ and $c_{0}=\rho_{0}(2\pi RT_{*})^{-3/2}$. 
We, thus, obtain the following inequality from (\ref{eq:kinetic}):
\begin{align}
  \frac{\partial}{\partial t}  \{\langle & f\ln\frac{f}{c_{0}}\rangle 
+\rho\ln(\frac{T^{3/2}}{T_{*}^{3/2}}\frac{\rho_{0}}{\rho})+\frac{\rho}{RT_{*}}(\mathcal{A}+\frac{1}{2}v^{2})\} \nonumber \\
&+\frac{\partial}{\partial X_{i}}\Big\{\langle\xi_{i}f\ln\frac{f}{c_{0}}\rangle+\rho v_{i}\ln(\frac{T^{3/2}}{T_{*}^{3/2}}\frac{\rho_{0}}{\rho})\nonumber \\
& +\frac{1}{RT_{*}}\{\rho(\mathcal{A}+\frac{1}{2}v^{2})v_{i}+\frac{1}{2}\langle c_{i}\bm{c}^{2}f\rangle+p_{ij}v_{j}\}\Big\}+\frac{\rho v_{i}}{RT_{*}}\frac{\partial\Phi_{L}}{\partial X_i}\le0,\label{eq:H1}
\end{align}
where the equality holds only when $f=\rho M_{*}$.

Now we integrate (\ref{eq:H1}) with respect to $\bm{X}$. 
After some lines of calculations with the aid of the mass balance equation,
we first note that
\begin{equation}
 \int_{D}\rho v_{i}\frac{\partial\Phi_{L}}{\partial X_i}d\bm{X} 
=  \int_{D}\frac{\partial\rho}{\partial t}\Phi_{L}d\bm{X}, \label{eq:equivalence}
\end{equation}
and that
\begin{equation}
 \int_{D}\frac{\partial\rho}{\partial t}\Phi_{L}d\bm{X} 
=  \frac{d}{dt}\int_{D}\rho\Phi_{L}d\bm{X}
  -\int_{D}\frac{\partial\rho}{\partial t}\Phi_{L}d\bm{X}, \label{eq:extforce}
\end{equation}
(see Appendix~\ref{sec:DSE}). Hence, we have
\begin{equation}
 \int_{D}\frac{\rho v_{i}}{RT_{*}}\frac{\partial\Phi_{L}}{\partial X_i}d\bm{X}
=\frac{1}{2RT_{*}}\frac{d}{dt}\int_{D}\rho\Phi_{L}d\bm{X}.\label{eq:simplification}
\end{equation}
With (\ref{eq:simplification})
in mind, we introduce the following quantities%
\begin{subequations}\label{free energies}
\begin{align}
\mathcal{F} =\langle f\ln\frac{f}{c_{0}}\rangle+&\rho\ln(\frac{T^{3/2}}{T_{*}^{3/2}}\frac{\rho_{0}}{\rho})+\frac{\rho}{RT_{*}}(\mathcal{A}+\frac{1}{2}\bm{v}^{2}+\frac{1}{2}\Phi_{L}),\\
\mathcal{F}_{i}  =\langle\xi_{i}f\ln\frac{f}{c_{0}}\rangle&+\rho v_{i}\ln(\frac{T^{3/2}}{T_{*}^{3/2}}\frac{\rho_{0}}{\rho})
\nonumber \\
&+\frac{1}{RT_{*}}\{\rho(\mathcal{A}+\frac{1}{2}v^{2})v_{i}+\frac{1}{2}\langle c_{i}\bm{c}^{2}f\rangle+p_{ij}v_{j}\}.
\end{align}
\end{subequations}%
By the substitution of the above into (\ref{eq:H1})
integrated over the spatial domain $D$, we have
\begin{equation}
\frac{d}{dt}\int_{D}\mathcal{F}d\bm{X}+\int_{D}\frac{\partial\mathcal{F}_{i}}{\partial X_{i}}d\bm{X}=\int_{D}\langle(\ln\frac{f}{\rho M_{*}})\,C_{*}[f]\rangle d\bm{X}\le0.\label{eq:Htheorem}
\end{equation}
Since the system is periodic, the second term on the left-hand side
vanishes because of the Gauss divergence theorem. Then, we are left
with
\begin{equation}
\frac{d}{dt}\mathcal{M}(t)=\frac{d}{dt}\int_{D}\mathcal{F}d\bm{X}=\int_{D}\langle(\ln\frac{f}{\rho M_{*}})\,C_{*}[f]\rangle d\bm{X}\le0,\label{eq:meanH}
\end{equation}
where $\mathcal{M}(t) \equiv\int_{D}\mathcal{F}d\bm{X}$, which is reduced to (see Appendix~\ref{sec:DSE})
\begin{equation}
\mathcal{M}(t) =\int_{D}\{\langle f\ln\frac{f}{\rho_{0}M_{*}}\rangle+\frac{1}{RT_{*}}\int\Phi_{S}d\rho+\frac{\rho}{2RT_{*}}\Phi_{L}[\rho]\}d\bm{X}.\label{eq:defM}
\end{equation}
This is the functional to be minimized in time. 

Note that the last equality in (\ref{eq:meanH}) holds only when $f=\rho M_{*}$.
Moreover, if $f$ is a local Maxwellian with temperature $T_{*}$,
then $\langle f\ln\frac{f}{c_{0}}\rangle+\rho\ln(\frac{T^{3/2}}{T_{*}^{3/2}}\frac{\rho_{0}}{\rho})$
vanishes, up to a constant multiple of $\rho$, and the functional
$\mathcal{M}$ corresponds to the Helmholtz free energy plus the potential
energy of the \textit{tail} part of long-range attractive potential
{[}see the first line of (66); note that $\bm{v}=0$ and
$T=T_{*}$, if $f$ is a local Maxwellian with temperature $T_{*}${]}.
The present observation is thermodynamically reasonable, because the
system is in \textit{contact} with the thermal bath with temperature
$T_{*}$ and the volume of domain $D$ is fixed. 

In the case $\Phi_{L}=-\kappa(\partial^2\rho/\partial X_i^2)$, 
the third term of (\ref{eq:defM}) is reduced to
\begin{equation}
  \int_{D}\frac{\rho\Phi_{L}}{2RT_{*}}d\bm{X}
=-\frac{\kappa}{2RT_{*}}\int_{D}\rho\frac{\partial^2\rho}{\partial X_i^2} d\bm{X}
= \frac{\kappa}{2RT_{*}}\int_{D}(\frac{\partial\rho}{\partial X_i})^{2}d\bm{X},
\end{equation}
so that $\mathcal{M}$ is expressed as 
\begin{equation}
\mathcal{M}(t)=\int_{D}\left(\langle f\ln\frac{f}{\rho_{0}M_{*}}\rangle+\frac{1}{RT_{*}}\int\Phi_{S}d\rho+\frac{\kappa}{2RT_{*}}(\frac{\partial\rho}{\partial X_i})^{2}\right)d\bm{X}.
\end{equation}
The last term in the above is often regarded as an energy of interface
in the literature.

\section{Dimensionless formulation}

Let us introduce the following notation:%
\begin{subequations}\label{factor}
\begin{align}
& t  =t_{*}\tilde{t},\ X_{i}=Lx_{i},\ \xi_{i}=(2RT_{*})^{1/2}\zeta_{i},\ \zeta=|\bm{\zeta}|,\ \rho=\rho_{0}\tilde{\rho},\\
&  f=\frac{\rho_{0}}{(2RT_{*})^{3/2}}\tilde{f}=c_{0}\pi^{3/2}\tilde{f},\ E=\pi^{-3/2}\exp(-\zeta^{2}),\ F_{i}=\frac{2RT_{*}}{L}\tilde{F}_{i},\\
&  \phi=2RT_{*}\tilde{\phi},\ \Phi_{S}=2RT_{*}\tilde{\Phi}_{S},\ \Phi_{L}=2RT_{*}\tilde{\Phi}_{L},\ \Psi=\frac{2RT_{*}}{(\rho_{0}/m)L^{3}}\tilde{\Psi},\\
&  \kappa=(2RT_{*}L^{2}/\rho_{0})\tilde{\kappa},\ a=\tilde{a}RT_{*}/\rho_{0},\ b=\tilde{b}/\rho_{0},\ A(\rho)=A_{0}\tilde{A}(\tilde{\rho}).
\end{align}
\end{subequations}%
The original equation is then reduced to
\begin{subequations}\label{modeldless}
\begin{align}
\mathrm{Sh}\frac{\partial\tilde{f}}{\partial\tilde{t}}
& +\zeta_{i}\frac{\partial\tilde{f}}{\partial x_{i}}+\tilde{F}_{i}\frac{\partial\tilde{f}}{\partial\zeta_{i}}=\frac{2}{\sqrt{\pi}}\frac{1}{\mathrm{Kn}}\tilde{C}_{*}[\tilde{f}],\label{eq:dimensionlessKM}\\
\tilde{C}_{*}&[\tilde{f}]=\tilde{A}(\tilde{\rho}E-\tilde{f}),\ \tilde{\rho}[\tilde{f}]=\!\!\int\tilde{f}d\bm{\zeta},\  E=\pi^{-3/2}\exp(-\zeta^{2}),\ \tilde{A}>0,\\
 \tilde{F}_{i}&=-\frac{\partial\tilde{\phi}}{\partial x_{i}},\quad\tilde{\phi}=\tilde{\Phi}_{S}(\tilde{\rho})+\tilde{\Phi}_{L},
\end{align}
\end{subequations}%
where
\begin{subequations}
\begin{align}
\tilde{\Phi}_{S} &=-\frac{1}{2}\ln(1-\tilde{b}\tilde{\rho})-\tilde{a}\tilde{\rho}+\frac{1}{2}\frac{\tilde{b}\tilde{\rho}}{1-\tilde{b}\tilde{\rho}},\\
\int\tilde{\Phi}_{S}d\tilde{\rho} &=-\frac{1}{2}\tilde{\rho}\ln(1-\tilde{b}\tilde{\rho})-\frac{1}{2}\tilde{a}\tilde{\rho}^{2},\\
\tilde{\Phi}_{L}(\bm{x}) &=\int_{\mathbb{R}^3}\tilde{\Psi}(|\tilde{\bm{r}}|)\{\tilde{\rho}(\bm{x}+\tilde{\bm{r}})-\tilde{\rho}(\bm{x})\}d\tilde{\bm{r}}\nonumber \\
\mathrm{or} & =-\tilde{\kappa}\Delta\tilde{\rho},\quad\mathrm{with}\quad\tilde{\kappa}=-\frac{1}{6}\int_{\mathbb{R}^3}\tilde{\Psi}(|\tilde{\bm{r}}|)\tilde{r}^{2}d\tilde{\bm{r}}=-\frac{2}{3}\pi\int_0^\infty\tilde{\Psi}(\tilde{r})\tilde{r}^{4}d\tilde{r},
\end{align}
and
\begin{equation}
\mathrm{Sh}=\frac{L}{t_{*}(2RT_{*})^{1/2}},\quad\mathrm{Kn}=\frac{(8RT_{*}/\pi)^{1/2}}{A_{0}L}.
\end{equation}
\end{subequations}%
Here and in what follows, $\Delta=\partial^{2}/\partial x_{i}^{2}$.
We also introduce the dimensionless quantities for the moments of
$f$ , i.e., $v_{i}=(2RT_{*})^{1/2}\tilde{v}_{i}$, $p=\rho_{0}RT_{*}\tilde{p}$,
$p_{ij}=\rho_{0}RT_{*}\tilde{p}_{ij}$, $e=RT_{*}\tilde{e}$, $s=R\tilde{s}$,
$\mathcal{A}=RT_{*}\tilde{\mathcal{A}}$, and $T=T_{*}\tilde{T}$.
Then, the quantities with tilde are expressed as%
\begin{subequations}\label{macrodless}
\begin{align}
&
\tilde{\rho}\tilde{v}_{i} =\langle\zeta_{i}\tilde{f}\rangle,\quad\tilde{T}=\frac{2}{3\tilde{\rho}}\langle\tilde{\bm{c}}^{2}\tilde{f}\rangle,\quad\tilde{p}=\frac{2}{3}\langle\tilde{\bm{c}}^{2}\tilde{f}\rangle+2\int\tilde{\rho}\tilde{\Phi}_{S}^{\prime}d\tilde{\rho}=\tilde{\rho}\tilde{T}+2\int\tilde{\rho}\tilde{\Phi}_{S}^{\prime}d\tilde{\rho},\\
&
\tilde{p}_{ij} =2\langle\tilde{c}_{i}\tilde{c}_{j}\tilde{f}\rangle+2\int\tilde{\rho}\tilde{\Phi}_{S}^{\prime}d\tilde{\rho}\delta_{ij},\quad\tilde{\rho}\tilde{e}=\langle\tilde{\bm{c}}^{2}\tilde{f}\rangle-\tilde{a}\tilde{\rho}^{2},\\
&
\tilde{\rho}\tilde{\mathcal{A}} =\tilde{\rho}(\tilde{e}-\tilde{s})
=\frac{3}{2}\tilde{\rho}+\tilde{\rho}\ln\tilde{\rho}+2\int\tilde{\Phi}_{S}d\tilde{\rho},\label{eq:localfree_energy}
\end{align}
\end{subequations}%
where $\tilde{\bm{c}}=(2RT_{*})^{-1/2}\bm{c}$.
Here and in what follows, $\langle\,\cdot\,\rangle=\int\,\cdot\,d\bm{\zeta}$.
In the meantime, the equation of state (\ref{eq:vdw}) is recast as
\begin{equation}
\tilde{p}=\frac{\tilde{\rho}\tilde{T}}{1-\tilde{b}\tilde{\rho}}-\tilde{a}\tilde{\rho}^{2}.\label{eq:vdw_dless}
\end{equation}

The balance laws of mass and momentum are rewritten as%
\begin{subequations}\label{massmomentum}
\begin{align}
&
 \mathrm{Sh}\frac{\partial \tilde{\rho}}{\partial\tilde{t}} 
+\frac{\partial}{\partial x_{i}}(\tilde{\rho}\tilde{v}_{i})=0,\\
&
\mathrm{Sh}\frac{\partial}{\partial \tilde{t}} (\tilde{\rho}\tilde{v}_{j})
+\frac{\partial}{\partial x_{i}}(\tilde{\rho}\tilde{v}_{i}\tilde{v}_{j}+\frac{1}{2}\tilde{p}_{ij})+\tilde{\rho}\frac{\partial\tilde{\Phi}_{L}}{\partial x_j}=-\frac{2}{\sqrt{\pi}}\frac{\tilde{A}}{\mathrm{Kn}}\tilde{\rho}\tilde{v}_{j}.
\end{align}
\end{subequations}%
Furthermore, by setting $\mathcal{F}=\rho_{0}\tilde{\mathcal{F}}$
and reminding $c_{0}=\rho_{0}(2\pi RT_{*})^{-3/2}$, we have
\begin{equation}
\tilde{\mathcal{F}}=\langle\tilde{f}\ln\frac{\tilde{f}}{E}\rangle+2\int\tilde{\Phi}_{S}d\tilde{\rho}+\tilde{\rho}\tilde{\Phi}_{L},
\end{equation}
and 
\begin{equation}
\frac{d\tilde{\mathcal{M}}}{d\tilde{t}}\le0,
\end{equation}
where $\tilde{\mathcal{M}}=(\rho_{0}L^{3})\mathcal{M}$ and it is
written as 
\begin{equation}
\tilde{\mathcal{M}}(\tilde{t})=\int_{\tilde{D}}\tilde{\mathcal{F}}d\bm{x}=\int_{\tilde{D}}\{\langle\tilde{f}\ln\frac{\tilde{f}}{E}\rangle+2\int\tilde{\Phi}_{S}d\tilde{\rho}+\tilde{\rho}\tilde{\Phi}_{L}\}d\bm{x},
\end{equation}
where $\tilde{D}$ is the dimensionless spatial region, the counterpart
of the dimensional one $D$. Remind that $\tilde{\mathcal{M}}$ is
non-increasing in time $\tilde{t}$ and reaches a stationary state
only when $\tilde{f}=\tilde{\rho}E$. 

When $\tilde{\Phi}_{L}=-\tilde{\kappa}\Delta\tilde{\rho}$, $\tilde{\mathcal{M}}$
is further reduced to%
\begin{subequations}
\begin{align}
\tilde{\mathcal{M}}(\tilde{t}) & =\int_{\tilde{D}}\{\langle\tilde{f}\ln\frac{\tilde{f}}{E}\rangle+2\int\tilde{\Phi}_{S}d\tilde{\rho}+\tilde{\kappa}(\frac{\partial\tilde{\rho}}{\partial x_i})^{2}\}d\bm{x},\\
 & 2\int\tilde{\Phi}_{S}d\tilde{\rho}=-\tilde{a}\tilde{\rho}^{2}-\tilde{\rho}\ln(1-\tilde{b}\tilde{\rho}),
\end{align}
because 
\begin{equation}
\int_{\tilde{D}}\tilde{\rho}\tilde{\Phi}_{L}d\bm{x}=-\tilde{\kappa}\int_{\tilde{D}}\tilde{\rho}\Delta\tilde{\rho}d\bm{x}=\tilde{\kappa}\int_{\tilde{D}}(\frac{\partial\tilde{\rho}}{\partial x_i})^{2}d\bm{x}.
\end{equation}
\end{subequations}

\section{Asymptotic analysis for small $\bm{\mathrm{Kn}}$\label{sec:Asymptotic-analysis-for}}

In the present section, we carry out the asymptotic analysis of (\ref{eq:dimensionlessKM})
for small $\mathrm{Kn}$, in order to study the behavior in the strong
interaction with the thermal bath. Hereafter, we drop tildes from
the dimensionless notation. Note that, if we set $A(\rho)=1$, the
nonlinearity comes solely from the self-consistent force field.

The original dimensionless equation (\ref{eq:dimensionlessKM}) recasts as%
\begin{subequations}
\begin{align}
\varepsilon\Big\{\mathrm{Sh}&\frac{\partial f}{\partial t} +\zeta_{i}\frac{\partial f}{\partial x_{i}}-\frac{\partial\phi}{\partial x_{i}}\frac{\partial f}{\partial\zeta_{i}}\Big\}=A(\rho)(\rho E-f),\label{eq:original}\\
&\phi=\Phi_{S}(\rho)+\Phi_{L}[\rho],\quad\rho[f]=\int fd\bm{\zeta},\quad E=\pi^{-3/2}\exp(-\zeta^{2}),
\end{align}
\end{subequations}%
where $\varepsilon=(\sqrt{\pi}/2)\mathrm{Kn}$.
When $\mathrm{Kn}$ or $\varepsilon$ is small, the right-hand side
is dominant in (\ref{eq:original}), and we are motivated to write
$f=f_{0}+g$, where $f_{0}=\rho E$. We construct $g$ by an iterative
procedure under the constraint $\int gd\bm{\zeta}=0$. From (\ref{eq:original}),
\begin{equation}
g=-\varepsilon(\mathrm{Sh}\frac{\partial f}{\partial t}+\zeta_{i}\frac{\partial f}{\partial x_{i}}-\frac{\partial\phi}{\partial x_{i}}\frac{\partial f}{\partial\zeta_{i}})\frac{1}{A(\rho)},\label{eq:g}
\end{equation}
and the constraint leads to
\begin{equation}
\mathrm{Sh}\frac{\partial\rho}{\partial t}+\frac{\partial}{\partial x_{i}}\int\zeta_{i}gd\bm{\zeta}=0.\label{eq:continuity}
\end{equation}
Our procedure below yields a successive approximation to $\int\zeta_{i}gd\bm{\zeta}$.

The first approximation $g_{1}$ is obtained by setting $f=f_{0}$
in (\ref{eq:g}), i.e.,
\begin{align}
g_{1}=& -\varepsilon\Big(\mathrm{Sh}\frac{\partial f_{0}}{\partial t}+\zeta_{i}\frac{\partial f_{0}}{\partial x_{i}}-\frac{\partial\phi}{\partial x_{i}}\frac{\partial f_{0}}{\partial\zeta_{i}}\Big)\frac{1}{A(\rho)} \nonumber \\
=& -\varepsilon\Big(\mathrm{Sh}\frac{\partial \rho}{\partial t}+\frac{\zeta_{i}}{\rho}\frac{\partial \rho}{\partial x_{i}}+2\zeta_{i}\frac{\partial\phi}{\partial x_{i}}\Big)\frac{\rho E}{A(\rho)}.
\end{align}
Because $g_{1}$ is an approximation to $g$ within the error of $o(\varepsilon)$,
it is enough that the constraint is satisfied within the same order of error,
namely $\int g_{1}d\bm{\zeta}=o(\varepsilon)$. Hence, by substitution
of the above expression of $g_{1}$, we see that $\varepsilon\mathrm{Sh}\partial\rho/\partial t=o(\varepsilon)$,
which implies $\mathrm{Sh}=o(1)$.%
\footnote{In the present analysis, the magnitude of $\mathrm{Sh}$ has not been assumed, except for that it is, at most, of $O(1)$. 
If we set $\mathrm{Sh}=O(1)$ at this stage, $\partial\rho/\partial t$ is of $o(1)$, which implies that the time scale in our dimensionless formulation is not proper to follow the time evolution for small $\varepsilon$. In this way, we find a proper size of $\mathrm{Sh}$ to be of $o(1)$.}%
 The first approximation is then
simply written as
\begin{equation}
g_{1}=-\varepsilon\zeta_{i}\Big(\frac{1}{\rho}\frac{\partial \rho}{\partial x_{i}}+2\frac{\partial\phi}{\partial x_{i}}\Big)\frac{\rho E}{A(\rho)}+o(\varepsilon),
\end{equation}
which yields
\begin{align}
\int\zeta_{i}g_{1}d\bm{\zeta}&= -\varepsilon\int\zeta_{i}\zeta_{j}\Big(\frac{1}{\rho}\frac{\partial \rho}{\partial x_{j}}+2\frac{\partial\phi}{\partial x_{j}}\Big)\frac{\rho E}{A(\rho)}d\bm{\zeta}+o(\varepsilon) \nonumber \\
&= -\varepsilon\frac{1}{3}\int\zeta^{2}\frac{\rho E}{A(\rho)}d\bm{\zeta}\Big(\frac{1}{\rho}\frac{\partial \rho}{\partial x_{i}}+2\frac{\partial\phi}{\partial x_{i}}\Big)+o(\varepsilon) \nonumber \\
&= -\varepsilon\frac{1}{2}\frac{\rho}{A(\rho)}\frac{\partial}{\partial x_{i}}(\ln\rho+2\phi)+o(\varepsilon). 
\end{align}
Therefore, the first approximation to (\ref{eq:continuity}) is given
by
\begin{equation}
\mathrm{Sh}\frac{\partial\rho}{\partial t}-\frac{\varepsilon}{2}\frac{\partial}{\partial x_{i}}\Big(\frac{\rho}{A(\rho)}\frac{\partial}{\partial x_{i}}(\ln\rho+2\phi)\Big)=o(\varepsilon).\label{eq:firstorder}
\end{equation}
To proceed to the second approximation, we set $f=f_{0}+g_{1}$ in (\ref{eq:g}). 
After some manipulations (see Appendix~\ref{sec:DSE}), we have
\begin{eqnarray}
g_{2}= -\varepsilon\frac{\rho}{A(\rho)}\frac{\partial}{\partial x_{i}}(\ln\rho+2\phi)\zeta_{i}E+\frac{\varepsilon^{2}}{A(\rho)}\Big\{\frac{\partial}{\partial x_{i}}\{\frac{\rho}{A(\rho)}\frac{\partial}{\partial x_{j}}(\ln\rho+2\phi)\}\nonumber \\
 \qquad+\frac{2\rho}{A(\rho)}\frac{\partial\phi}{\partial x_{i}}\frac{\partial}{\partial x_{j}}(\ln\rho+2\phi)\Big\}(\zeta_{i}\zeta_{j}-\frac{1}{2}\delta_{ij})E+o(\varepsilon^{2}). \label{eq:g_2nd}
\end{eqnarray}
It is seen that the above form has already satisfied the
constraint $\int g_{2}d\bm{\zeta}=o(\varepsilon^{2})$. Therefore,
by substitution, the second approximation to (\ref{eq:continuity})
is obtained as
\begin{equation}
\mathrm{Sh}\frac{\partial\rho}{\partial t}-\frac{\varepsilon}{2}\frac{\partial}{\partial x_{i}}\Big\{\frac{\rho}{A(\rho)}\frac{\partial}{\partial x_{i}}(\ln\rho+2\phi)\Big\}=o(\varepsilon^{2}).\label{secondorder}
\end{equation}
It should be noted that the accuracy estimate of (\ref{secondorder}) is improved
by one order from the stage of (\ref{eq:firstorder}), although the resulting equation
looks the same.

Further reduction of (\ref{secondorder}) is possible by using the concrete form of $\phi$.
Since $\phi=\Phi_{S}(\rho)+\Phi_{L}$, we have 
\begin{equation}
\mathrm{Sh}\frac{\partial\rho}{\partial t}-\frac{\varepsilon}{2}\frac{\partial}{\partial x_{j}}\{\frac{1}{A(\rho)}(1+2\rho\Phi_{S}^{\prime})\frac{\partial\rho}{\partial x_{j}}+\frac{2\rho}{A(\rho)}\frac{\partial\Phi_{L}}{\partial x_{j}}\}=o(\varepsilon^{2}).
\end{equation}
By setting $\mathrm{Sh}=\varepsilon$ and taking the limit $\varepsilon\to0$,
we have
\begin{equation}
\frac{\partial\rho}{\partial t}-\frac{\partial}{\partial x_{j}}\{\frac{1}{A(\rho)}(\frac{1}{2}+\rho\Phi_{S}^{\prime})\frac{\partial\rho}{\partial x_{j}}+\frac{\rho}{A(\rho)}\frac{\partial\Phi_{L}}{\partial x_{j}}\}=0.\label{eq:density}
\end{equation}
Remind that%
\begin{subequations}\label{dlP}
\begin{align}
\Phi_{S}(\rho)  =&-\frac{1}{2}\ln(1-b\rho)-a\rho+\frac{1}{2}\frac{b\rho}{1-b\rho},\\
\Phi_{S}^{\prime}(\rho)  =&-a+\frac{1}{2}\frac{b(2-b\rho)}{(1-b\rho)^{2}},\\
\Phi_{L}[\rho](\bm{x}) & =\int_{\mathbb{R}^3}\Psi(|\bm{r}|)\{\rho(\bm{x}+\bm{r})-\rho(\bm{x})\}d\bm{r}\nonumber \\
 \mathrm{or} & =-\kappa\Delta\rho,\quad\mathrm{with}\quad\kappa=-\frac{1}{6}\int_{\mathbb{R}^3}\Psi(|\bm{r}|)r^{2}d\bm{r}.
\end{align}
\end{subequations} 

For later convenience, let us introduce a rescaled density $\chi=b\rho$
and rewrite (\ref{eq:density}) for the case that $\Phi_{L}$ is local.
Then, we have%
\begin{subequations}\label{rch}
\begin{align}
\frac{\partial\chi}{\partial t}&- \frac{\partial}{\partial x_{j}}\{\frac{\chi}{A(\chi/b)}\frac{\partial}{\partial x_{j}}(\Phi-K\frac{\partial^{2}\chi}{\partial x_{i}^{2}})\}=0, \\
& \Phi=-c\chi+\frac{1}{2}\frac{1}{(1-\chi)}+\frac{1}{2}\ln\frac{\chi}{1-\chi},\quad K=\frac{\kappa}{b},\quad c=\frac{a}{b},
\end{align}
\end{subequations}%
where $0<\chi<1$ and $\Phi(\chi)$ is related
to $\Phi_{S}$ as $\Phi=(1/2)\ln\chi-1/2+\Phi_{S}(\chi/b)$. By setting
$A\equiv1$, we have a following Cahn\textendash Hilliard type equation:%
\begin{subequations}\label{Cahn}
\begin{align}
\frac{\partial\chi}{\partial t}-& \frac{\partial}{\partial x_{j}}\{\chi\frac{\partial}{\partial x_{j}}(\Phi-K\frac{\partial^{2}\chi}{\partial x_{i}^{2}})\}=0,\\
& \Phi=-c\chi+\frac{1}{2}\frac{1}{(1-\chi)}+\frac{1}{2}\ln\frac{\chi}{1-\chi}.
\end{align}
\end{subequations}

\subsection{Linear stability of a uniform state \label{subsec:Linear-stability}}

In the present subsection, we study the linear stability of the uniform
state on the basis of (\ref{Cahn}). Substituting $\chi=\chi_{\mathrm{av}}+\epsilon\exp(\sigma t+ik_{j}x_{j})$
and retaining the terms of $O(\epsilon)$,%
\footnote{If we set $\rho_0$ as the average density, then 
$\chi_\mathrm{av}$ is identical to $b$ occurring in \eqref{rch}.}
 we obtain%
\begin{subequations}
\begin{align}
\sigma & =\chi_{\mathrm{av}}\{-\Phi^{\prime}(\chi_{\mathrm{av}})-Kk^{2}\}k^{2},\\
 & \Phi^{\prime}(\chi)=-c+\frac{1}{2}\frac{1}{\chi(1-\chi)^{2}},\quad k^{2}\equiv k_{i}^{2}.
\end{align}
\end{subequations}%
Thus, $\sigma$ is positive when $Kk^{2}+\Phi^{\prime}(\chi_{\mathrm{av}})<0$.
Namely, when $c>\frac{1}{2}\frac{1}{\chi_{\mathrm{av}}(1-\chi_{\mathrm{av}})^{2}}$,
the uniform state $\chi=\chi_{\mathrm{av}}$ is (linear) unstable.
The most rapidly growing mode $k_{\mathrm{mr}}$ can be found by the
condition $d\sigma/dk=0$, which leads to $\{-2\Phi^{\prime}(\chi_{\mathrm{av}})-4Kk_{\mathrm{mr}}^{2}\}k_{\mathrm{mr}}=0$,
namely
\begin{equation}
k_{\mathrm{mr}}^{2}=-\frac{1}{2K}\Phi^{\prime}(\chi_{\mathrm{av}})=\frac{1}{2K}\{c-\frac{1}{2}\frac{1}{\chi_{\mathrm{av}}(1-\chi_{\mathrm{av}})^{2}}\}.
\end{equation}

\subsection{Free energy at a local equilibrium and stationary states\label{subsec:Another-viewpoint-based}}

Let us recall the functional $\mathcal{M}$ for the case $\Phi_{L}=-\kappa\Delta\rho$:%
\begin{subequations}
\begin{align}
\mathcal{M}(t) & =\int_{D}\{\langle f\ln\frac{f}{E}\rangle+2\int\Phi_{S}d\rho+\kappa(\frac{\partial\tilde{\rho}}{\partial x_i})^{2}\}d\bm{x},\label{eq:CHfree1}\\
 & 2\int\Phi_{S}d\rho=-a\rho^{2}-\rho\ln(1-b\rho).
\end{align}
\end{subequations}%
Under the assumption $f=\rho E$, $\mathcal{M}$
is reduced to 
\begin{equation}
\mathcal{M}(t)=\int_{D}(\rho\ln\rho+2\int\Phi_{S}d\rho-\kappa\rho\Delta\rho)d\bm{x}.\label{eq:CHfree2}
\end{equation}
Note that, except for a constant multiple of $\rho$, the sum of the
first two terms of the integrand in (\ref{eq:CHfree2}) is identical
with $\rho\mathcal{A}$ {[}see (\ref{eq:localfree_energy}){]}. It
is identical with $(2/b)\int\Phi(\chi)d\chi$ as well, except for
a constant multiple of $\chi$. We therefore simply call $\int\Phi d\chi$
a \textit{local free energy} in the sequel. A similar result for the
nonlocal self-consistent force field can be found, e.g., in \cite{RW02}
and \cite{CCELM05}. We rewrite (56) in terms of the rescaled
density $\chi$ to have an equivalent functional
\begin{equation}
\mathcal{M_{\chi}}(t) \equiv b\mathcal{M}(t)
=\int_{D}\{\chi(\ln\frac{\chi}{1-\chi}-c\chi)-K\chi\Delta\chi-\chi\ln b\}d\bm{x}.
\label{MX}
\end{equation}
Here $\ln b$ in the integrand plays the same role as the Lagrangian
multiplier under the constraint $\int_{D}\chi d\bm{x}=\mathrm{const.}$
and is to be written as $\lambda$ below. We can find stationary states
by the variational method, namely by the condition that $\delta\mathcal{M}_{\chi}/\delta\chi=0$,
which yields
\begin{equation}
\int_{D}\{(\ln\frac{\chi}{1-\chi}-c\chi)+\chi(\frac{1}{\chi}+\frac{1}{1-\chi}-c)-2K\Delta\chi-\lambda\}\delta\chi d\bm{x}=0.
\end{equation}
Therefore,
\begin{figure}
\centering
\begin{tabular}{ccc}
\includegraphics[bb=78bp 530bp 442bp 779bp,clip,width=0.4\textwidth]{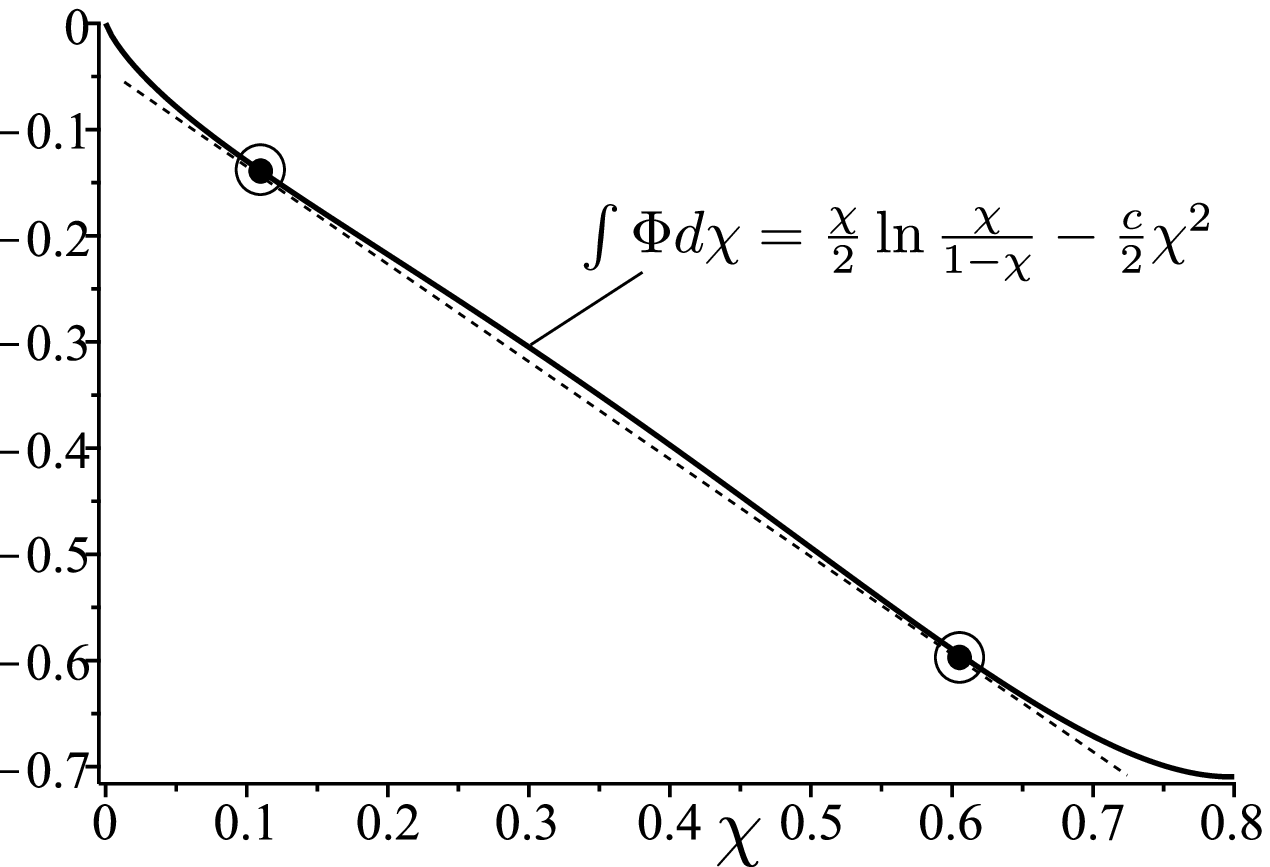}&&
\includegraphics[bb=0bp 0bp 365bp 250bp,clip,width=0.4\textwidth]{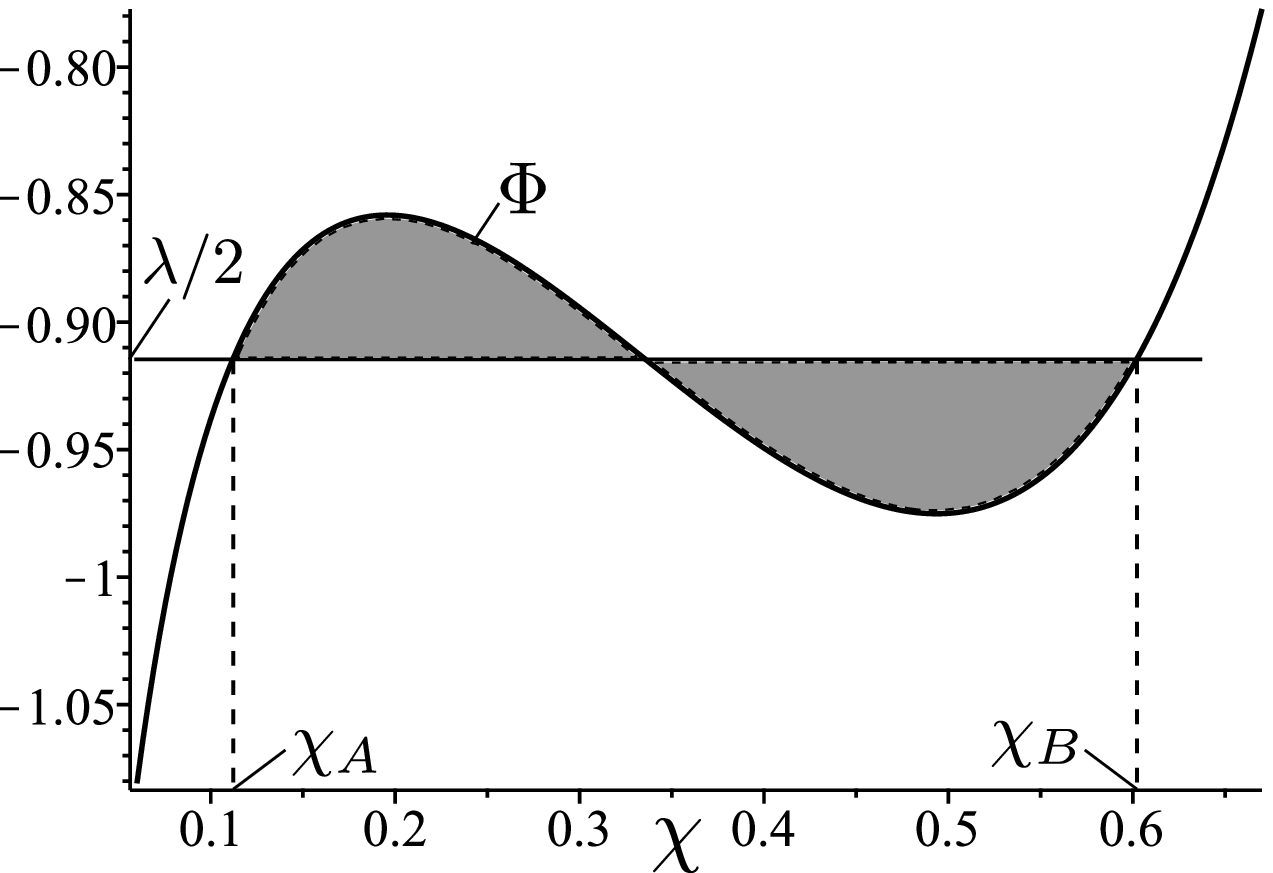}\\
\small (a)  && \small (b) 
\end{tabular}
\caption{Local free energy and two coexisting states. (a) The \textit{local
free energy} $\int\Phi d\chi$ with $c=3.95$ and the two-points tangential
line (dashed line) that determines the coexisting states. (b) The
derivative of the \textit{local free energy} $\Phi$ and the equi-area
rule for determining the coexisting states, $\chi_{A}$ and $\chi_{B}$.
\label{fig: The-local-free}}
\end{figure}
\begin{figure}
\centering
\begin{tabular}{ccc}
\includegraphics[bb=0bp 15bp 432bp 288bp,clip,width=0.45\textwidth]{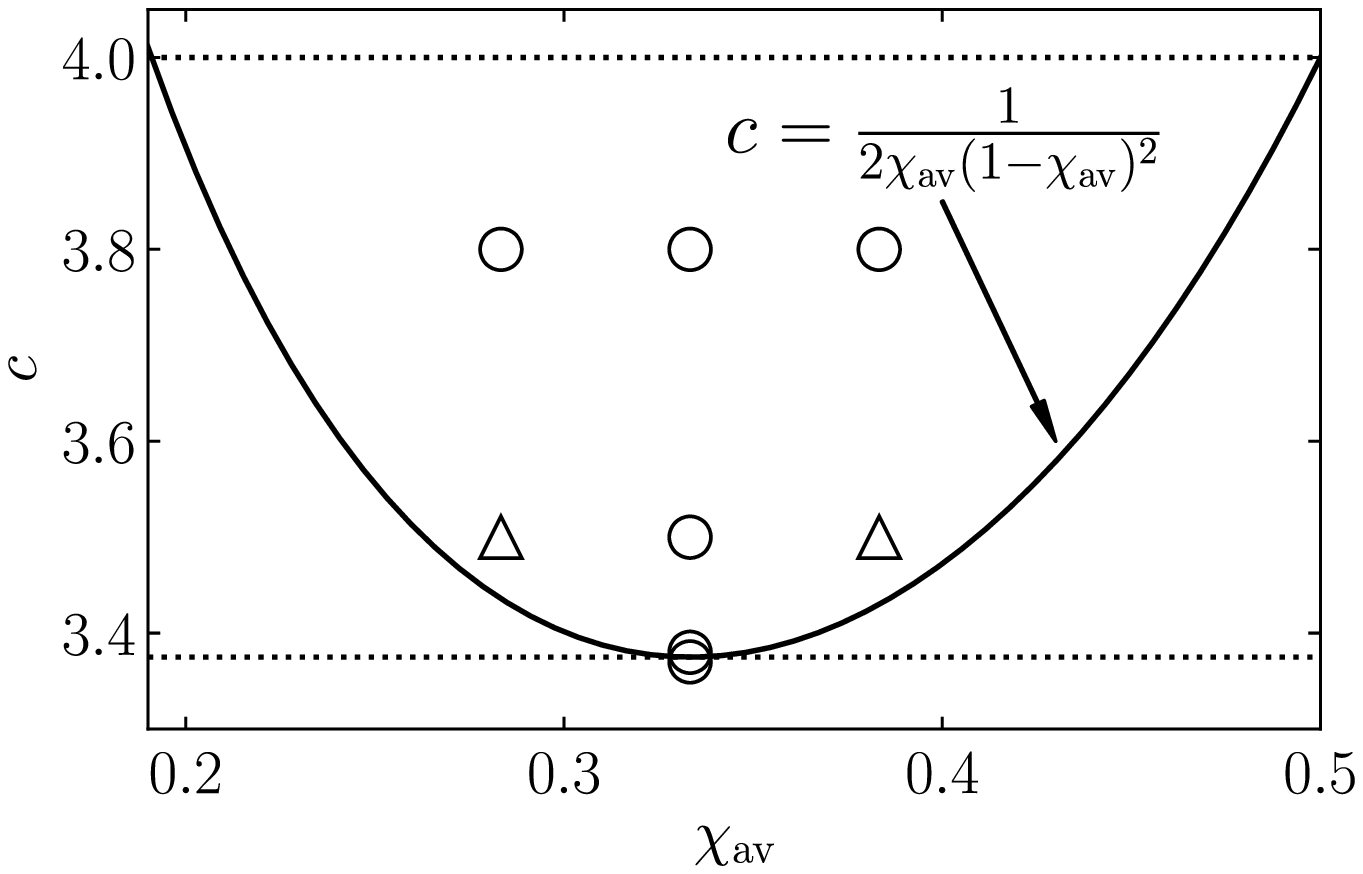}&&
\includegraphics[bb=0bp 15bp 432bp 288bp,clip,width=0.45\textwidth]{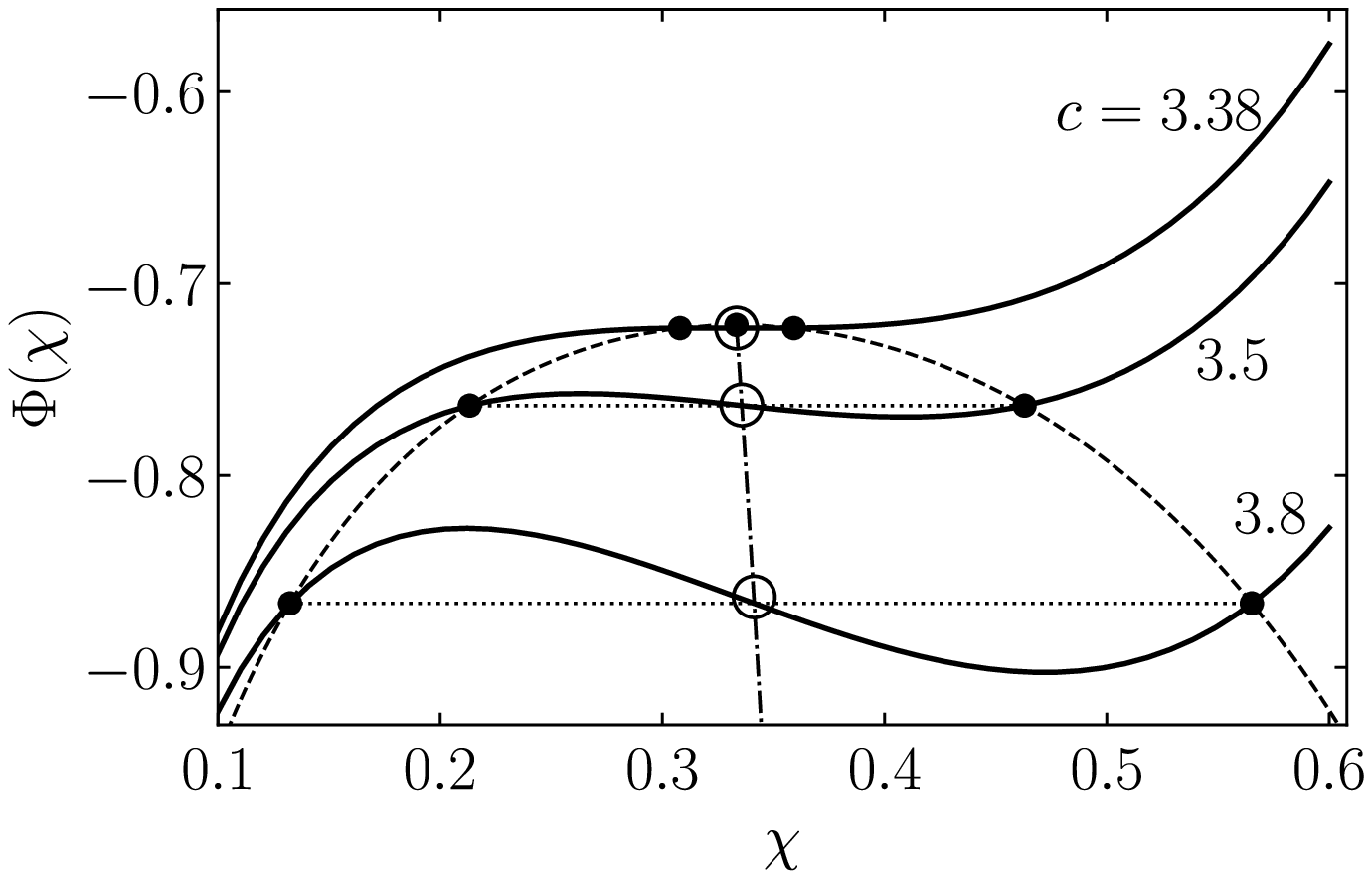}\\
\small (a)  && \small (b) 
\end{tabular}
\caption{The neutral curve of linear stability and $\Phi(\chi)$ for different
values of $c$. (a) The neutral curve (solid line) and the parameters
of numerical simulations (open circles and triangles). (b)
$\Phi(\chi)$ for $c=3.38$, $3.5$, and $3.8$. In (a), a uniform
state is linearly unstable in the region above the neutral curve.
In (b), the three points that are determined by the equi-area rule
are indicated with a pair of closed circles and an open circle. The
dashed and dot\textendash dash lines are, respectively, the locus
of the former and that of the latter in changing the value of $c$.\label{fig:diagram}}
\end{figure}
\begin{eqnarray}
K\Delta\chi && =\frac{1}{2}\ln\frac{\chi}{1-\chi}+\frac{1}{2}\frac{1}{1-\chi}-c\chi-\frac{1}{2}\lambda=\Phi(\chi)-\frac{1}{2}\lambda.
\end{eqnarray}
In one dimensional case, the above equation can be interpreted as
a motion of point mass in a potential field $-\int\Phi(\chi)d\chi+(\lambda/2)\chi$
($\chi$, $x$, and $K$ are interpreted as the position, time, and
mass, respectively). This interpretation and following discussions
in the present paragraph are due to van Kampen \cite{VK64}. Let us
denote by $\chi_{A}$ and $\chi_{B}$ the values of $\chi$ at which
a local maximum of the potential is achieved and thus the identity
$\Phi(\chi_{A})=\Phi(\chi_{B})=\lambda/2$ ought to hold. This means
that there is a common tangential line of $\int\Phi d\chi$ to $\chi=\chi_{A}$
and $\chi=\chi_{B}$, the slope of which is $\lambda/2$ {[}see figure~\ref{fig: The-local-free}(a){]}.
In the meantime, from a mechanical point of view, the potential height
there should be the same in order for a spontaneous transition from
one to the other to occur. Therefore $-\int^{\chi_{A}}\Phi(\chi)d\chi+(\lambda/2)\chi_{A}=-\int^{\chi_{B}}\Phi(\chi)d\chi+(\lambda/2)\chi_{B}$
or $\int_{\chi_{A}}^{\chi_{B}}\Phi d\chi=(\lambda/2)(\chi_{B}-\chi_{A})$.
This implies that two shaded areas in figure~\ref{fig: The-local-free}(b)
are the same (equi-area rule). Both interpretations, namely the common
tangential line and the equi-area rule, often appear in the literature.

\begin{figure}
\centering
\begin{tabular}{cc}
\includegraphics[bb=30bp 0bp 355bp 278bp,clip,height=0.3\textwidth]{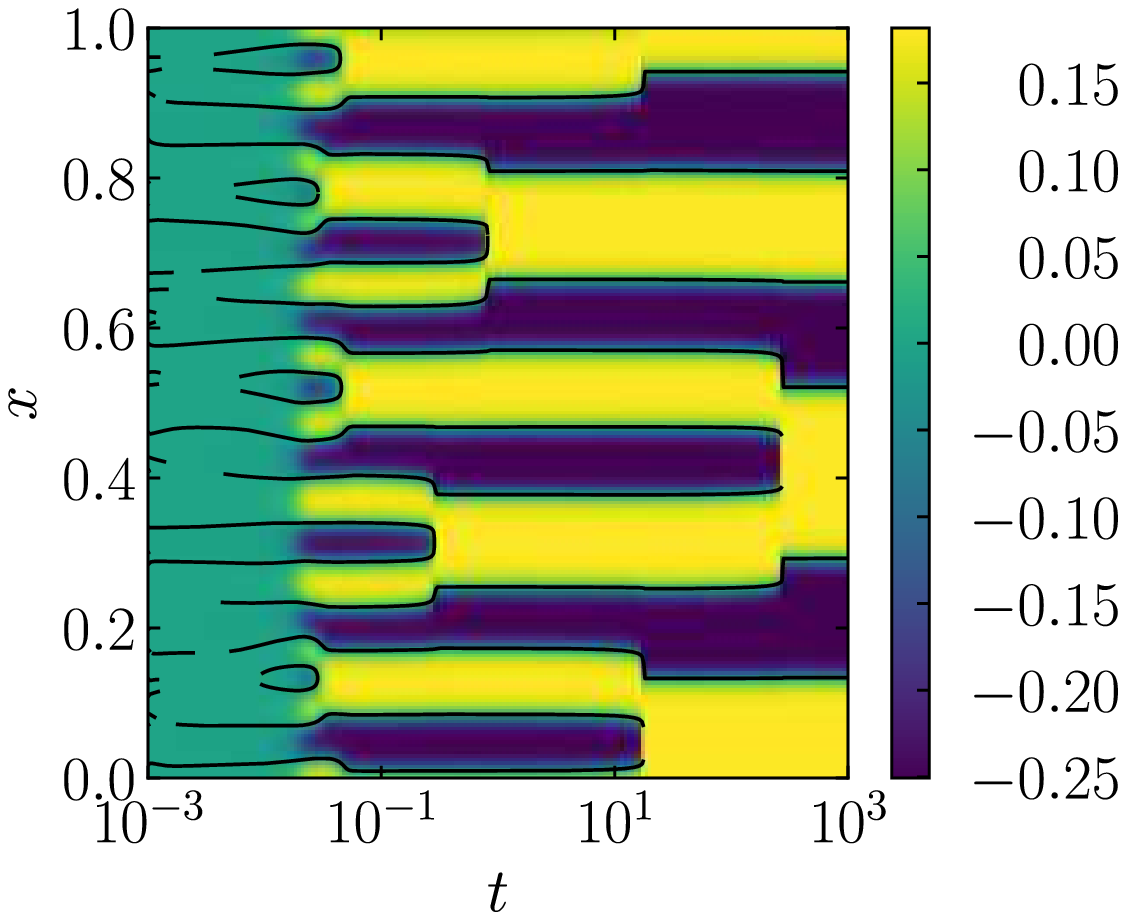}&
\includegraphics[bb=30bp 0bp 285bp 278bp,clip,height=0.3\textwidth]{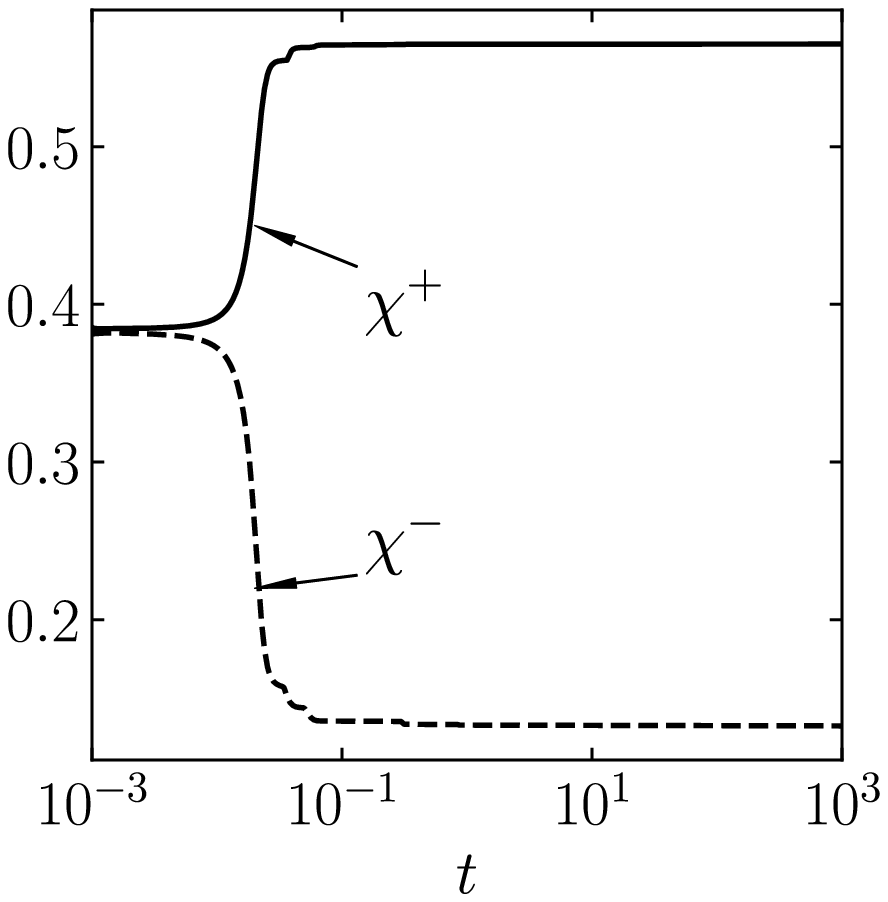}\\ 
\multicolumn{2}{c}{\small (a) $c=3.8$, $\chi_\mathrm{av}=23/60(=0.383)$}\\
&\\
\includegraphics[bb=30bp 0bp 355bp 278bp,clip,height=0.3\textwidth]{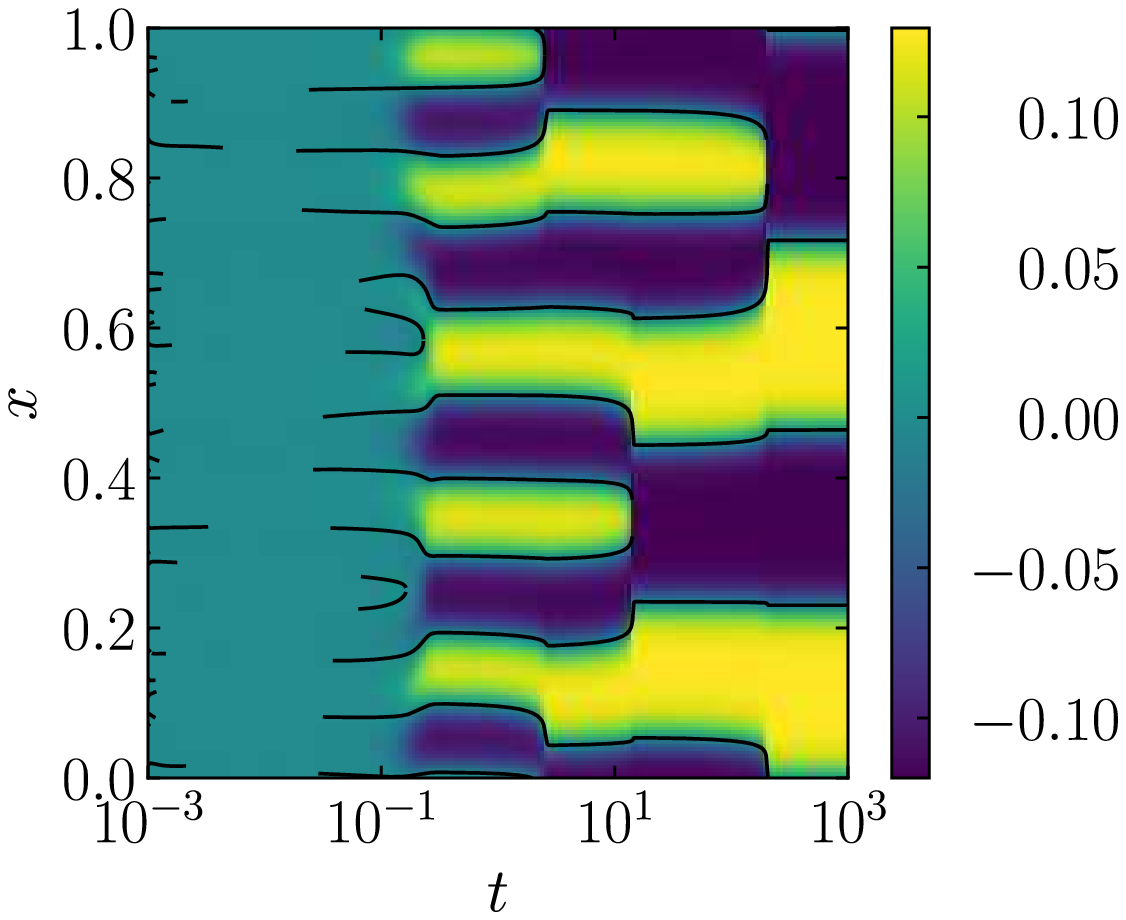}&
\includegraphics[bb=15bp 0bp 285bp 278bp,clip,height=0.3\textwidth]{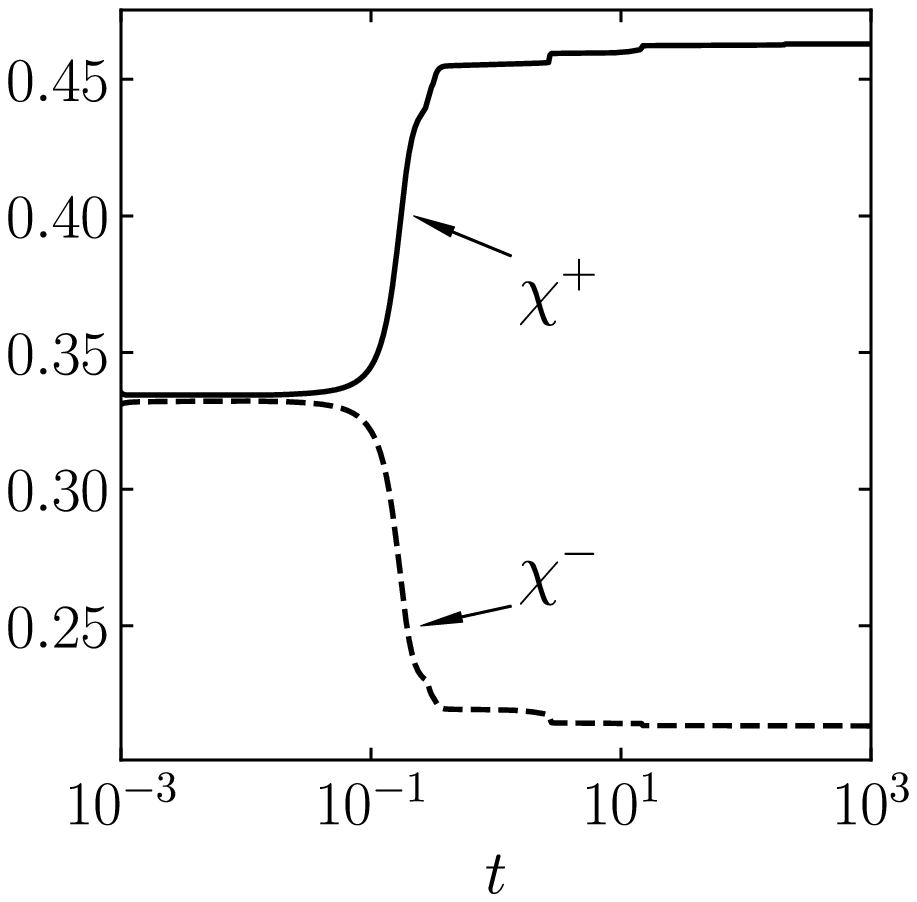}\\ 
\multicolumn{2}{c}{\small (b) $c=3.5$, $\chi_\mathrm{av}=1/3$}\\ 
&\\
\includegraphics[bb=30bp 0bp 355bp 278bp,clip,height=0.3\textwidth]{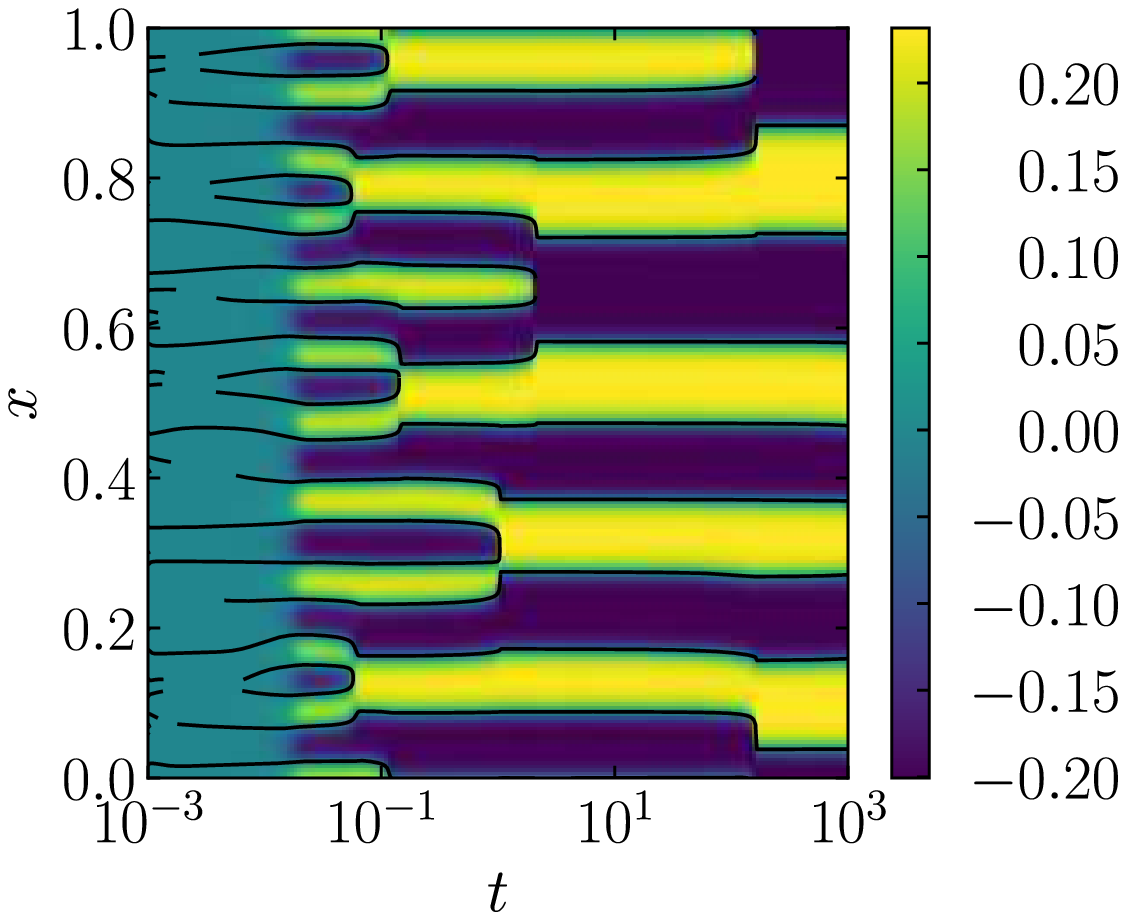}&
\includegraphics[bb=30bp 0bp 285bp 278bp,clip,height=0.3\textwidth]{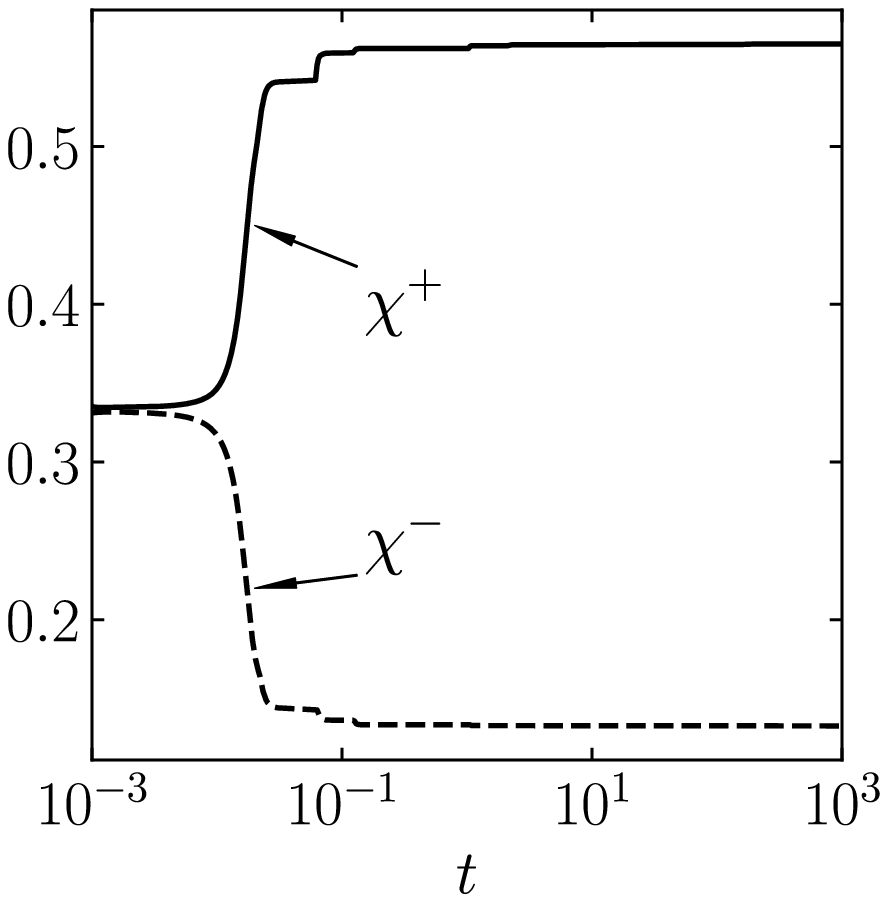}\\
\multicolumn{2}{c}{\small (c) $c=3.8$, $\chi_\mathrm{av}=1/3$}\\
\end{tabular}

\caption{Phase transition towards two coexisting phases for various sets of
parameters $(c,\chi_{\mathrm{av}})$, where $K$ is commonly set as
$4.3976\times10^{-5}$. In each case, the left panel shows the time
evolution of dense (light-colored) and dilute (dark-colored) phases.
The scale in the legend shows the deviation of $\chi$ from the average
$\chi_{\mathrm{av}}$. The contour line of $\chi=\chi_{\mathrm{\mathrm{av}}}$
is drawn as well, but it is omitted where $|\partial\chi/\partial x|<0.05$
for the clarity of figure. The right panel shows the time evolution
of the maximum/minimum values of $\chi$, say $\chi^{+}$ and $\chi^{-}$.
From (a) to (e), the values of $k_{\mathrm{mr}}/2\pi$ are 10.3, 6.00,
11.1, 1.20, and 10.2. The corresponding number of regions is observed
at the initiation of phase transition. The case (f) is in the range
of linear stability and thus neither $k_{\mathrm{mr}}$ nor phase
transition is found.\label{fig:C-H1D}}
\end{figure}

\begin{figure}
\centering
\begin{tabular}{cc}
\includegraphics[bb=30bp 0bp 355bp 278bp,clip,height=0.3\textwidth]{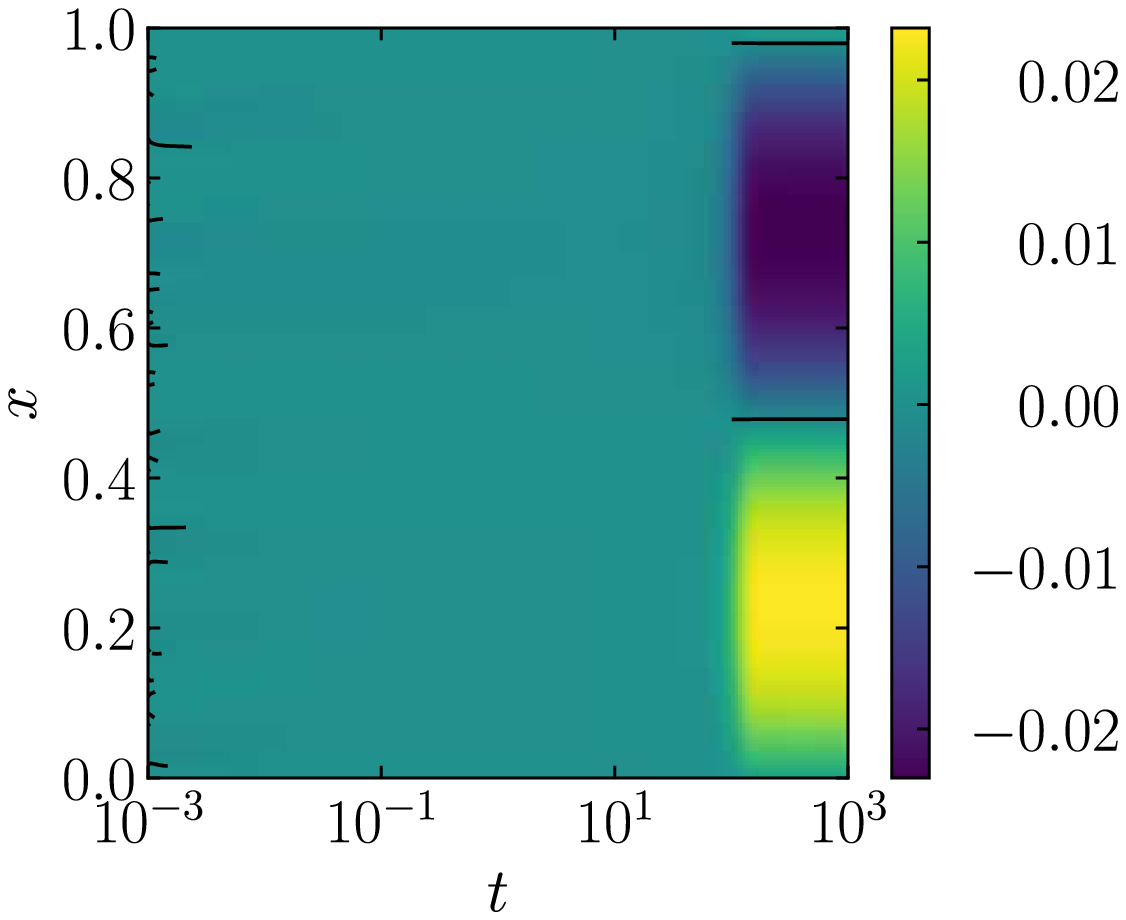}&
\includegraphics[bb=15bp 0bp 285bp 278bp,clip,height=0.3\textwidth]{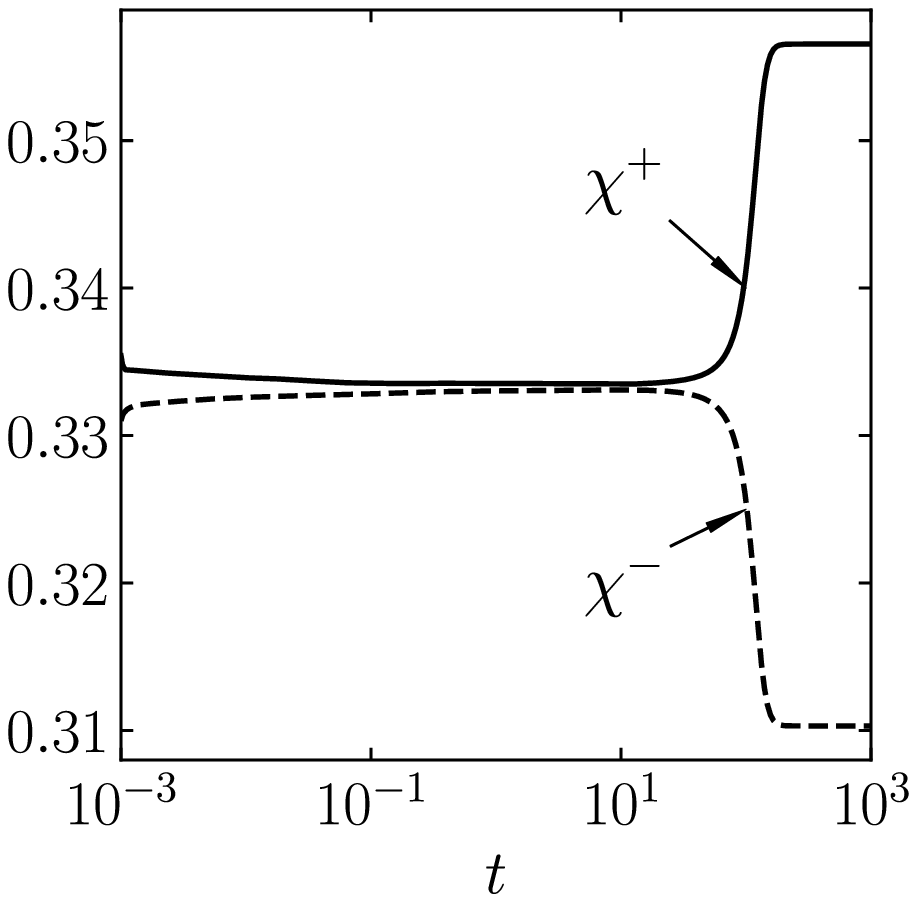}\\ 
\multicolumn{2}{c}{\small (d) $c=3.38$, $\chi_\mathrm{av}=1/3$}\\
&\\
\includegraphics[bb=30bp 0bp 355bp 278bp,clip,height=0.3\textwidth]{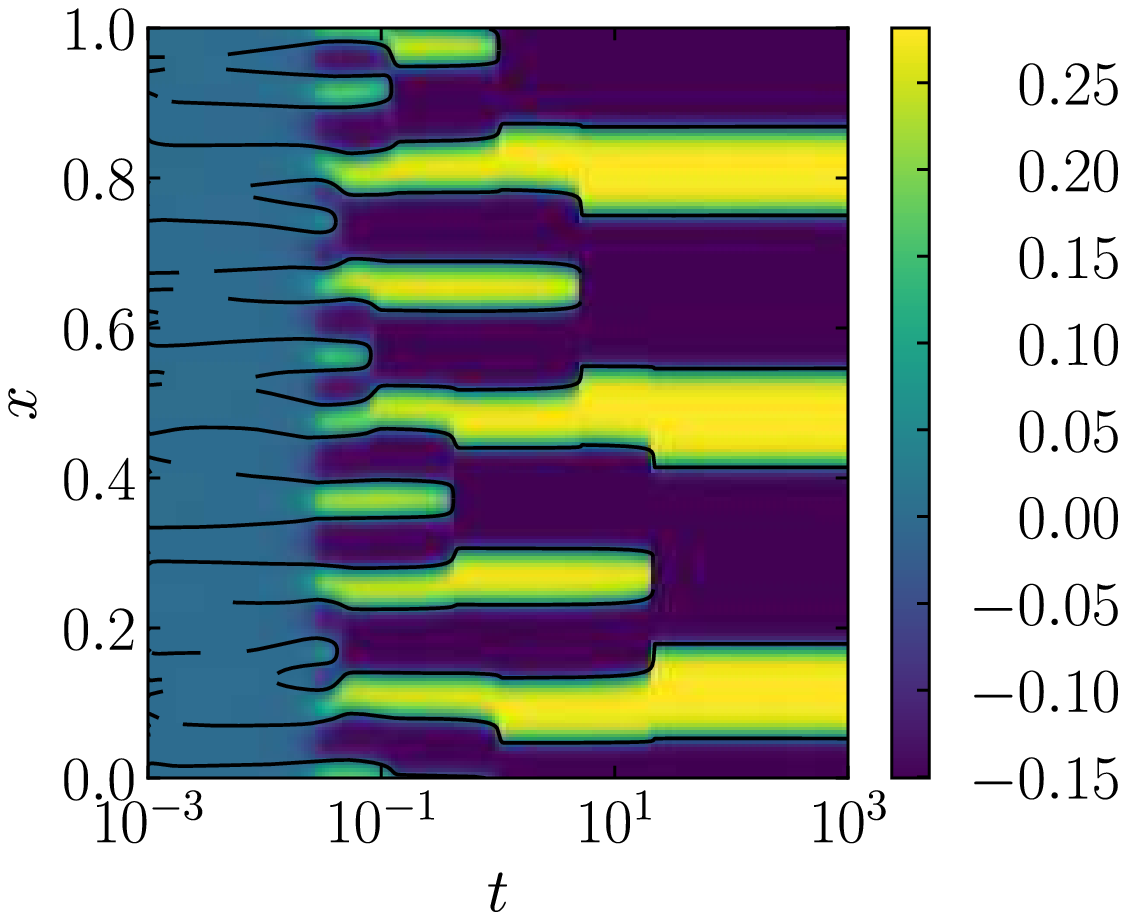}&
\includegraphics[bb=33bp 0bp 285bp 278bp,clip,height=0.3\textwidth]{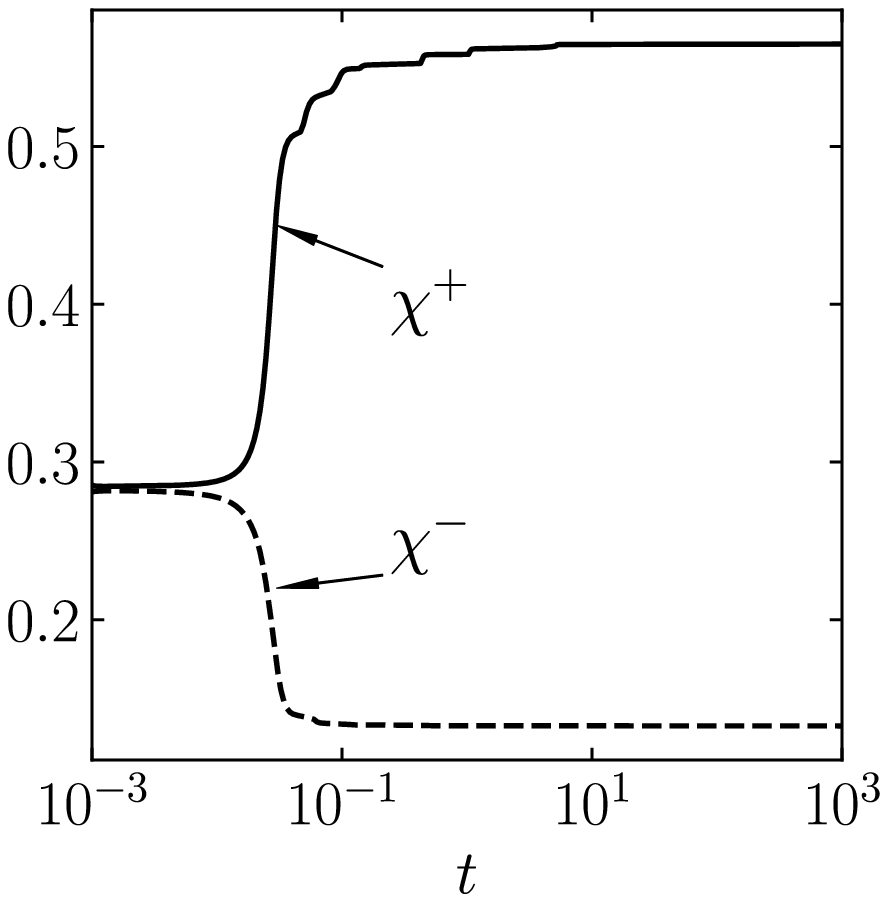}\\
\multicolumn{2}{c}{\small (e) $c=3.8$, $\chi_\mathrm{av}=17/60(=0.283)$}\\
&\\
\includegraphics[bb=30bp 0bp 355bp 278bp,clip,height=0.3\textwidth]{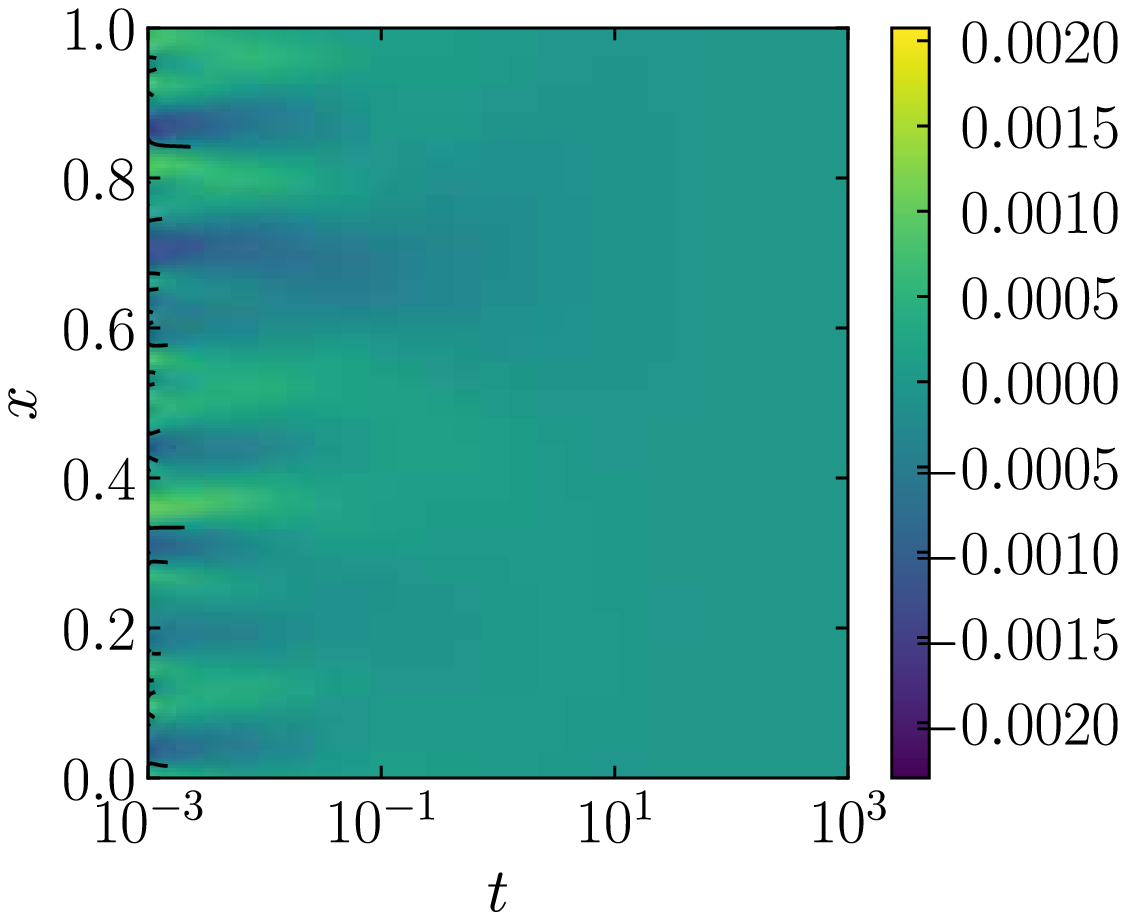}&
\includegraphics[bb=15bp 0bp 285bp 278bp,clip,height=0.3\textwidth]{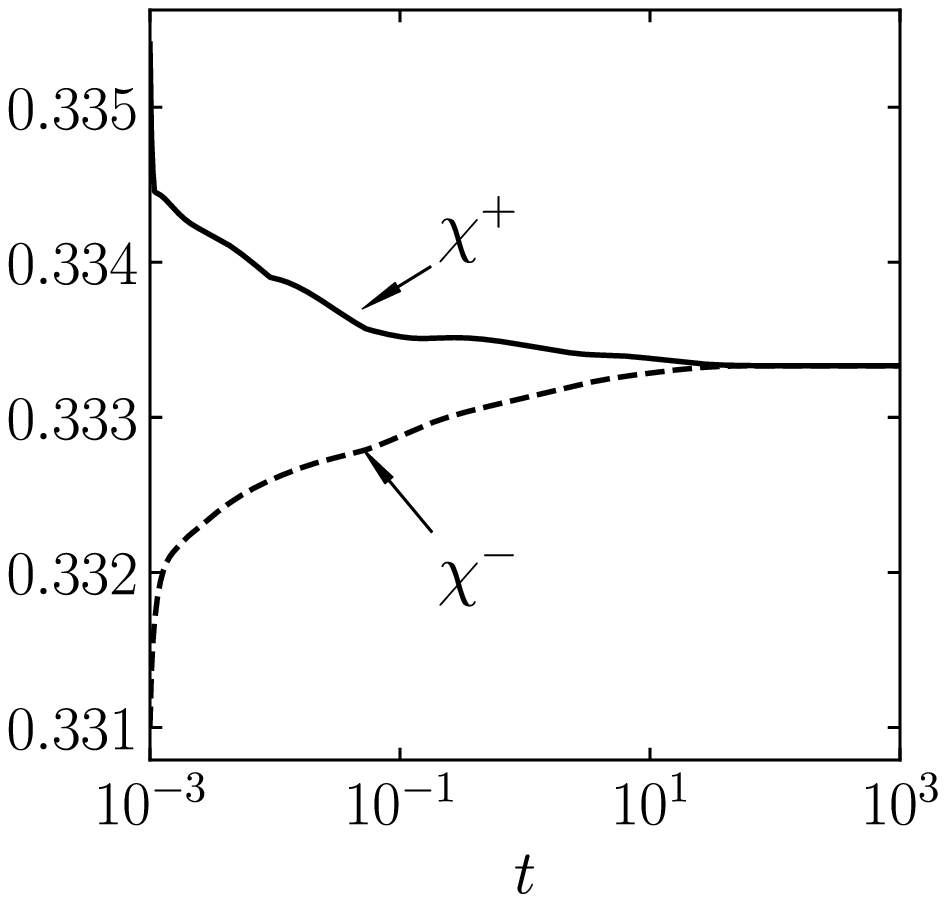}\\ 
\multicolumn{2}{c}{\small (f) $c=3.37$, $\chi_\mathrm{av}=1/3$}\\
\end{tabular}

\addtocounter{figure}{-1}
\caption{(continued from the previous page)}
\end{figure}

We now seek the condition that such different states can be found
based on the present shape of the function $\Phi(\chi)$. Because
\begin{equation}
\Phi^{\prime} =\frac{1}{2}\frac{1}{\chi(1-\chi)^{2}}-c\equiv\frac{1}{2}\frac{1}{h(\chi)}-c,
\end{equation}
and $h(\chi)$ takes its maximum $h_{\max}=4/27$ at $\chi=1/3$ and
a common minimum $h_{\min}=0$ at $\chi=0,\ 1$. Hence, the condition
$\Phi^{\prime}=0$ can be realized only when $c>27/8$. Furthermore,
$c<4$ should be satisfied in order for the van der Waals equation
of state (\ref{eq:vdw_dless}) with $T=1$ to assure the positive
pressure $p$ for any value of $\chi$. Therefore, we shall mainly
study the case $27/8<c<4$ in the sequel.

\subsection{Numerical simulations of the Cahn\textendash Hilliard type equation\label{subsec:Numerical-simulation-of}}

\begin{figure}
\centering\begin{tabular}{rrr}
\includegraphics[bb=0bp 0bp 324bp 268bp,clip,height=0.27\textwidth]{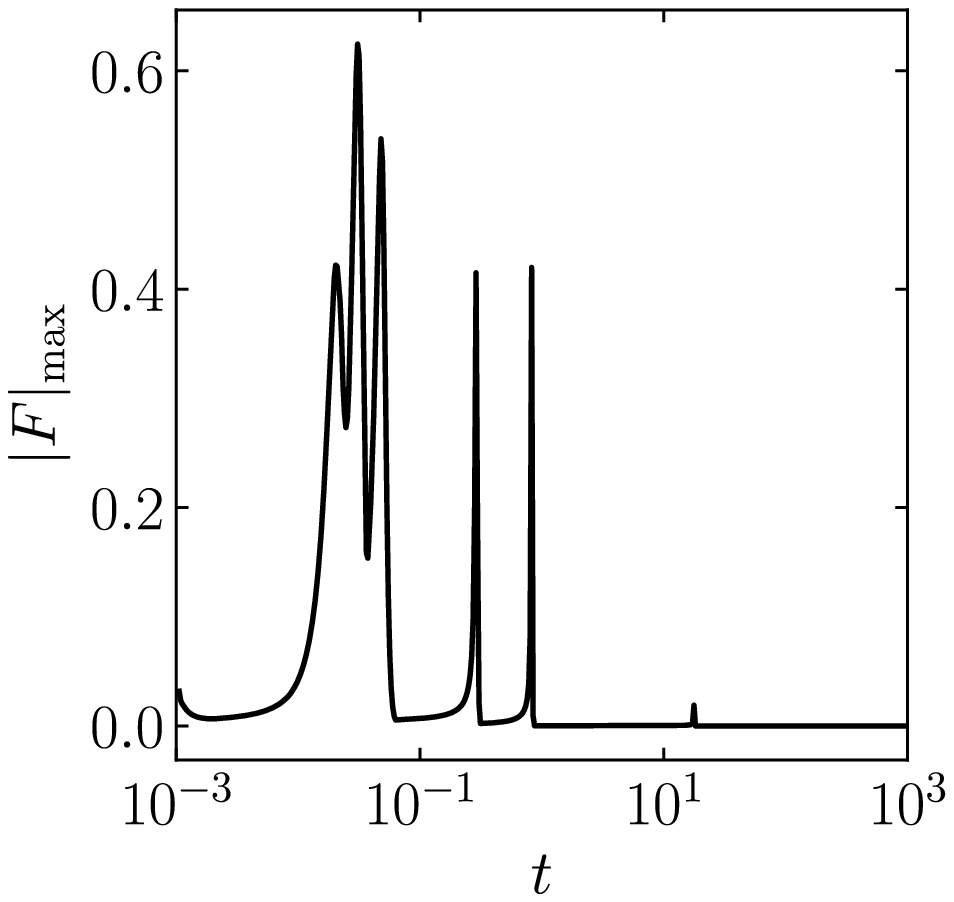}&
\includegraphics[bb=0bp 0bp 360bp 268bp,clip,height=0.27\textwidth]{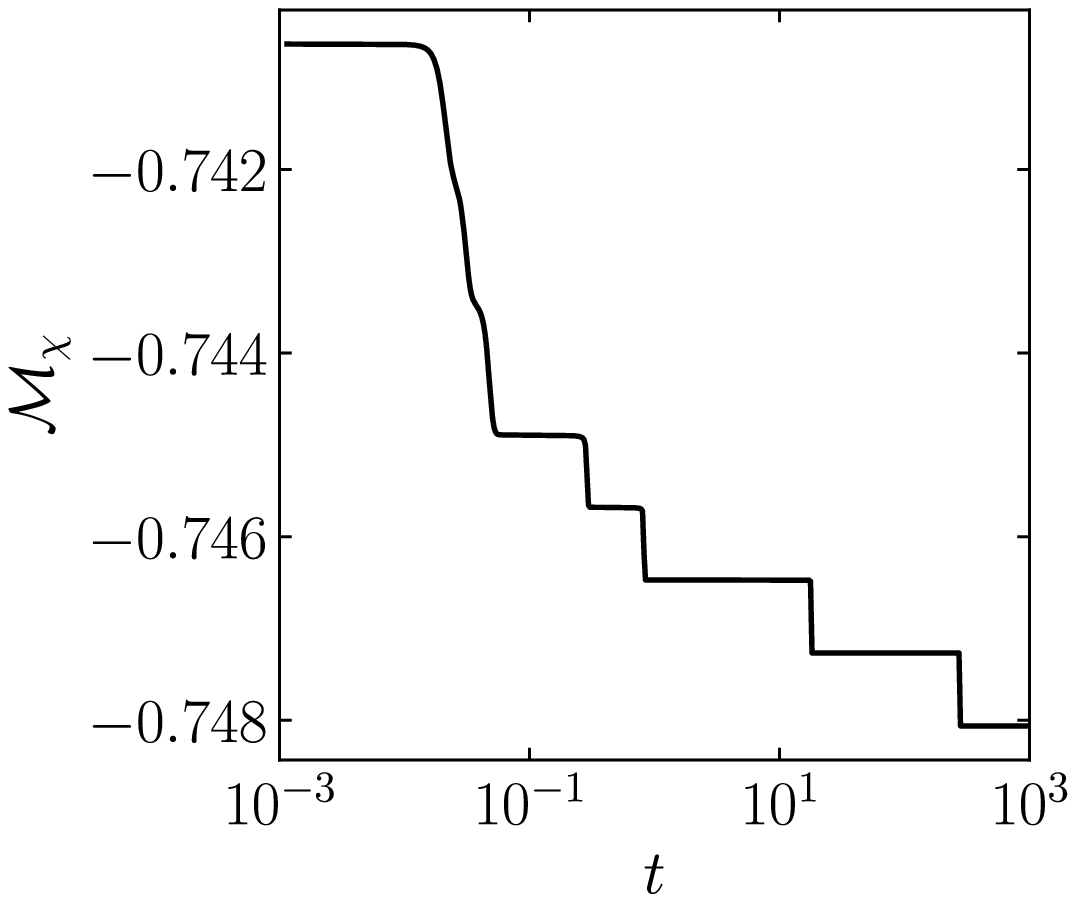}\\ 
\multicolumn{2}{c}{\small (a) $c=3.8$, $\chi_\mathrm{av}=23/60(=0.383)$}\\
& \\
&&\\
\includegraphics[bb=0bp 0bp 324bp 268bp,clip,height=0.27\textwidth]{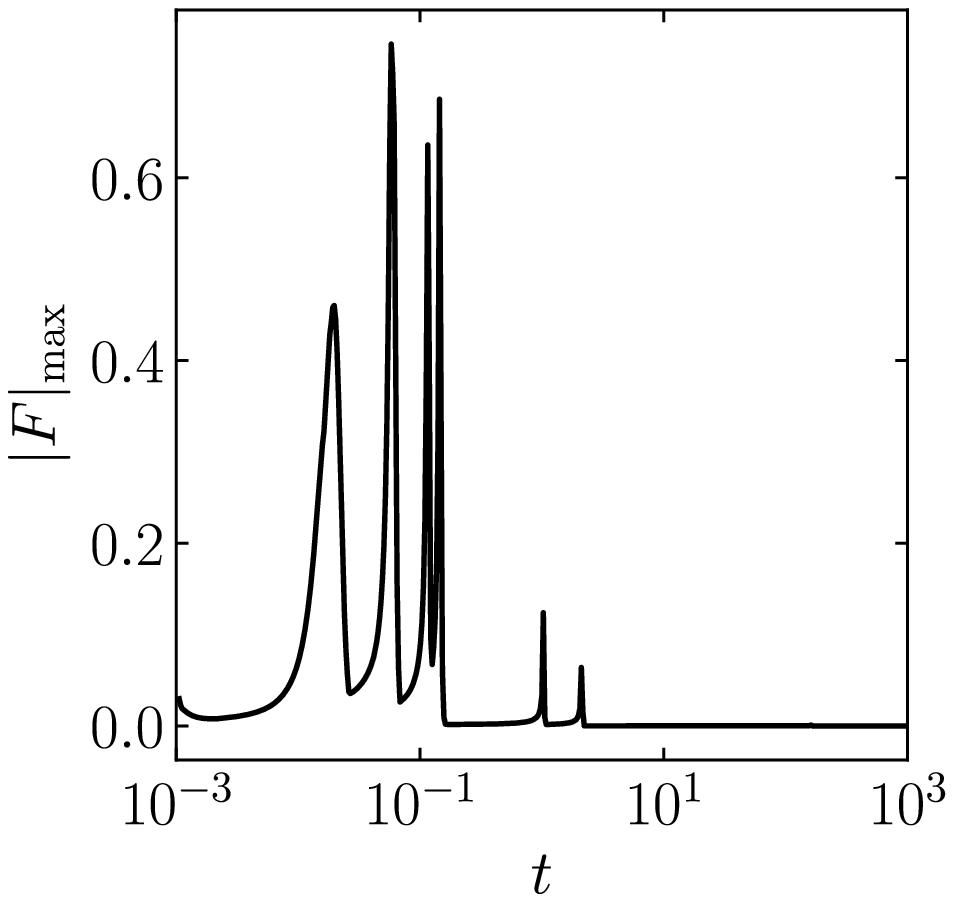}&
\includegraphics[bb=0bp 0bp 360bp 268bp,clip,height=0.27\textwidth]{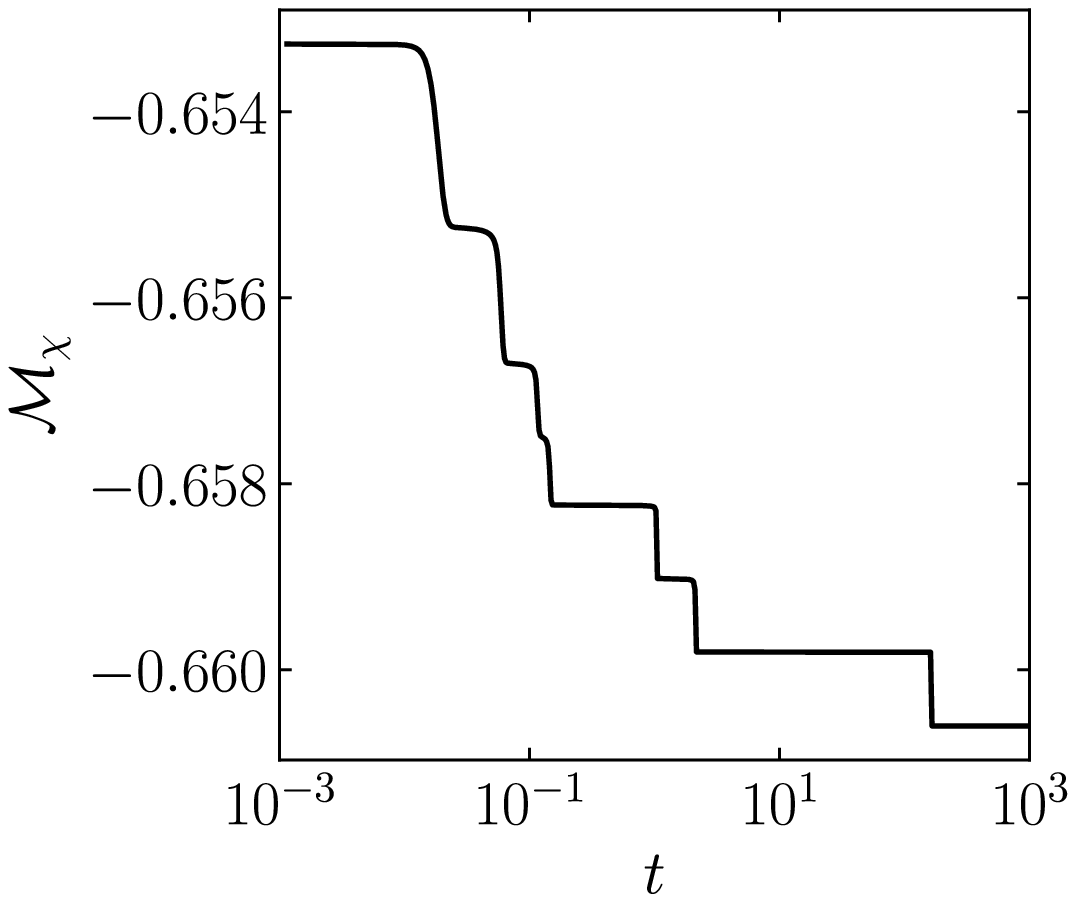}\\ 
\multicolumn{2}{c}{\small (b) $c=3.8$, $\chi_\mathrm{av}=1/3$}\\
& \\
&&\\
\includegraphics[bb=0bp 0bp 324bp 268bp,clip,height=0.27\textwidth]{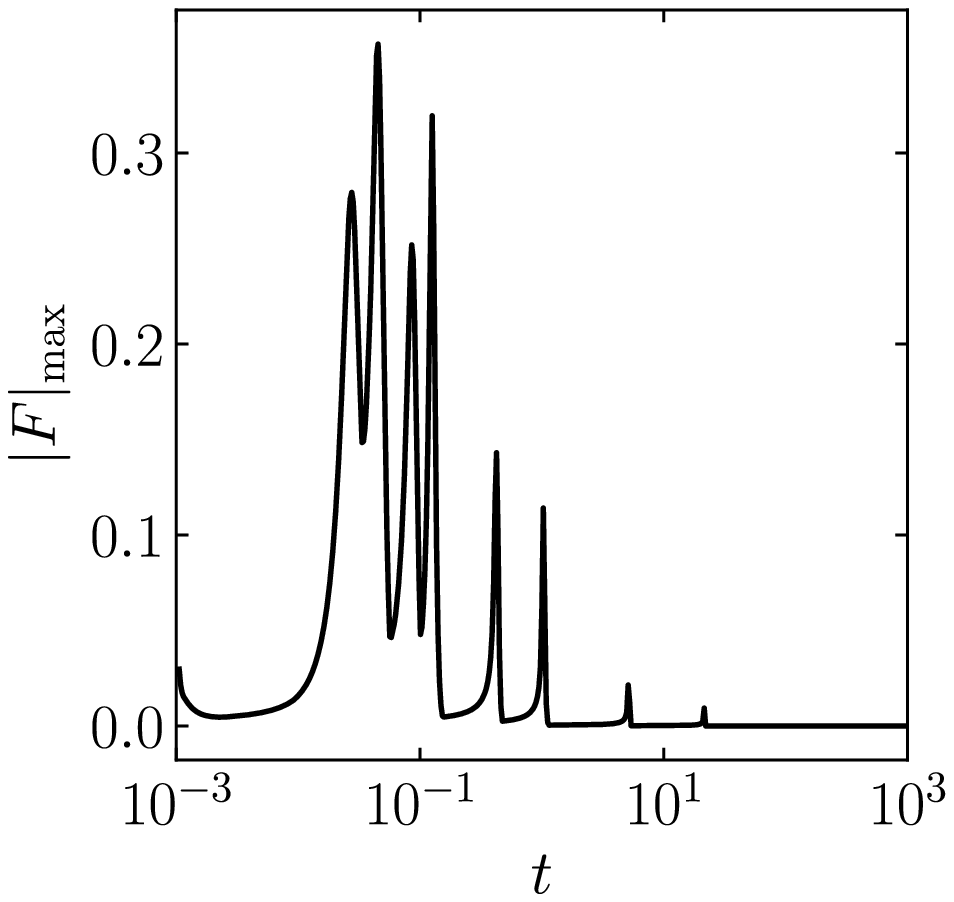}&
\includegraphics[bb=0bp 0bp 360bp 268bp,clip,height=0.28\textwidth]{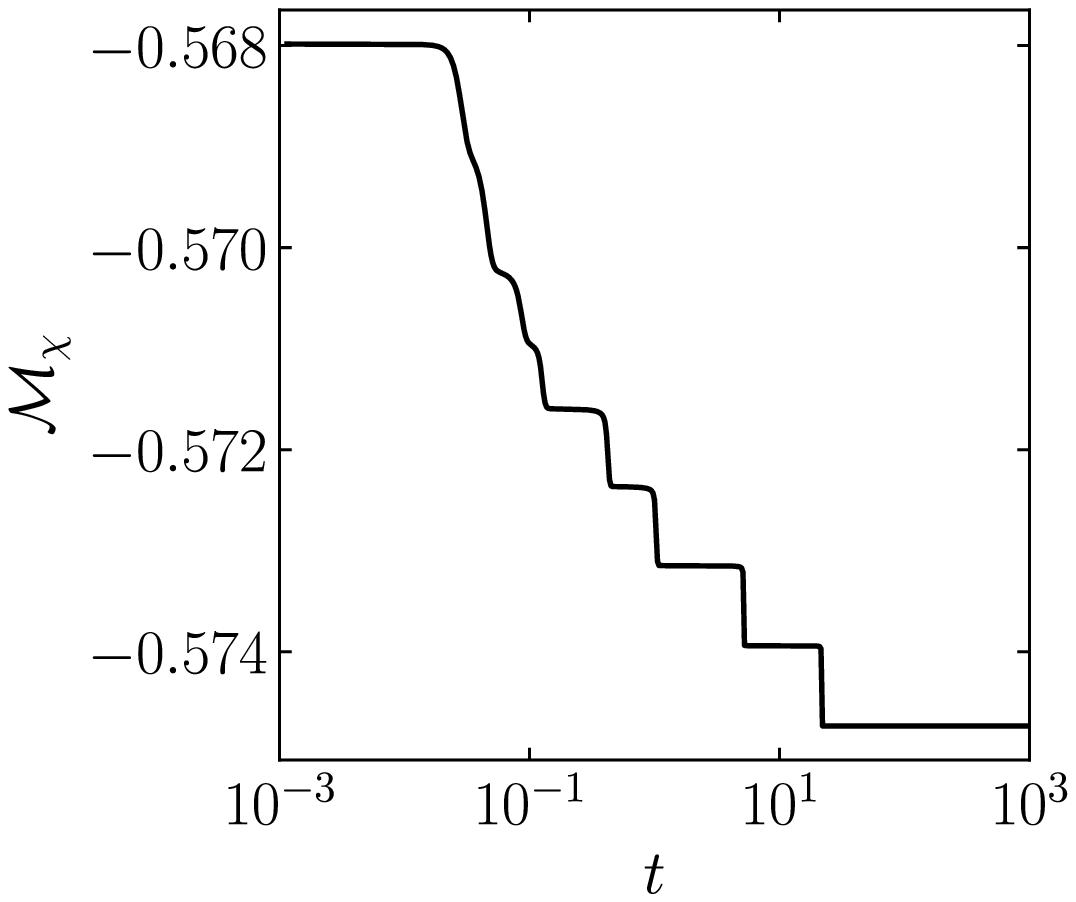}\\ 
\multicolumn{2}{c}{\small (c) $c=3.8$, $\chi_\mathrm{av}=17/60(=0.283)$}
\end{tabular}
\caption{Time evolution of the maximum mass flux $|F|_{\max}$ and the functional $\mathcal{M}_{\chi}$.\label{fig:flux}}
\end{figure}
We carried out numerical simulations of the Cahn\textendash Hilliard
type equation (\ref{Cahn}) for one-dimensional and two-dimensional
cases for different parameter pairs of $c$ and $\chi_{\mathrm{av}}$.
The chosen pairs are indicated by symbols in figure~\ref{fig:diagram}(a).
For the parameter pairs indicated by open triangles, the results of
two-dimensional simulations are just preliminary and will not be mentioned
in the sequel. In all the simulations, another parameter $K$ is commonly
set as $K=4.3976\times10^{-5}$ and the uniform state with $\chi=\chi_{\mathrm{av}}$
is initially disturbed by a Gaussian random noise with the standard
deviation of 0.001 (Further details of the initial disturbance can
be found in Appendix~\ref{sec:Numerical-scheme}). The value
of $K$ is chosen so that the most rapidly growing mode $k_{\mathrm{mr}}$
is about $6\times2\pi$ in the case $(c,\chi_{\mathrm{av}})=(3.5,1/3)$. 

We first show a part of the simulation results of one-dimensional
simulations in figure~\ref{fig:C-H1D}. In each simulation, $\mathcal{M_{\chi}}$
was monitored,%
\footnote{Here and in what follows, the contribution from the last term in \eqref{MX} is dropped
from the monitored value of $\mathcal{M}_\chi$, because it is constant under the present constraint.}
 together with the maximum mass flux, i.e.,
\begin{equation}
|F|_{\max}=\max_{x\in D}|F|,\quad F\equiv-\chi\frac{\partial}{\partial x}(\Phi-K\frac{\partial^{2}\chi}{\partial x^{2}}).\label{eq:F}
\end{equation}
Figure~\ref{fig:flux} shows the monitored results. In section~\ref{subsec:Another-viewpoint-based},
$\mathcal{M}_{\chi}$ has been evaluated under the assumption of the
local equilibrium state $f=\rho E$. The assumption is, however, broken
in the region where the mass flux is appreciable, as is clear in the
analysis in section~\ref{sec:Asymptotic-analysis-for}. In spite of
this discrepancy, the results show the monotonic decrease of $\mathcal{M}_{\chi}$,
which is consistent with the prediction in section~\ref{subsec:Another-viewpoint-based}.
The resulting consistency can be understood if we recompute $\mathcal{M}$
(or $\mathcal{M}_{\chi}$) with a better approximation of $f$ , i.e.,
\begin{equation}
f=\rho E\{1-\frac{\varepsilon\zeta_{i}}{A(\rho)}\Big(\frac{1}{\rho}\frac{\partial\rho}{\partial x_i}+2\frac{\partial\phi}{\partial x_{i}}\Big)\}+o(\varepsilon).
\end{equation}
Even with the refined $f$, we have
\begin{align}
& \langle f\ln\frac{f}{E}\rangle \nonumber\\
& \simeq\langle\rho E\{1-\frac{\varepsilon\zeta_{i}}{A(\rho)}\frac{\partial}{\partial x_i}\Big(\ln\rho+2\phi\Big)\}\ln[\rho\{1-\frac{\varepsilon\zeta_{i}}{A(\rho)}\frac{\partial}{\partial x_i}\Big(\ln\rho+2\phi\Big)\}]\rangle+o(\varepsilon) \nonumber \displaybreak[0] \\
&
 \simeq\langle\rho E\{1-\frac{\varepsilon\zeta_{i}}{A(\rho)}\frac{\partial}{\partial x_i}\Big(\ln\rho+2\phi\Big)\}\{\ln\rho-\frac{\varepsilon\zeta_{i}}{A(\rho)}\frac{\partial}{\partial x_i}\Big(\ln\rho+2\phi\Big)\}\rangle+o(\varepsilon) \nonumber \displaybreak[0] \\
&
 \simeq\langle\rho E\{1-\frac{\varepsilon\zeta_{i}}{A(\rho)}\frac{\partial}{\partial x_i}\Big(\ln\rho+2\phi\Big)\}\ln\rho-\rho E\frac{\varepsilon\zeta_{i}}{A(\rho)}\frac{\partial}{\partial x_i}\Big(\ln\rho+2\phi\Big)\rangle+o(\varepsilon) \nonumber \displaybreak[0] \\
&
 \simeq\langle\rho E\ln\rho\rangle+o(\varepsilon)=\rho\ln\rho+o(\varepsilon).
\end{align}
Thus, $\mathcal{M}$ (or $\mathcal{M}_{\chi}$) remains unchanged
up to $o(\varepsilon)$. Therefore, the deviation from the local Maxwellian
$f=\rho E$, which mainly occurs at the interface, does not affect
the minimization dynamics up to $o(\varepsilon)$. We therefore
regard $\mathcal{M}_{\chi}$ as a functional to be \textit{minimized} in time 
as well in the rest of the present subsection.

Now let us observe the results in figure~\ref{fig:flux} more closely.
The above form of $F$ in (\ref{eq:F}) suggests that the flux is
appreciable only at the interface. It is, however, appreciable only
in more limited situations, namely the initiation of phase transition
and subsequent emerging events of the same phases. Indeed, comparisons
with the corresponding cases in figure~\ref{fig:C-H1D} show a pulsive
response of $|F|_{\mathrm{max}}$ to those limited situations. The
functional $\mathcal{M}_{\chi}$ decreases monotonically, mostly with
stepwise falls that synchronize the pulsive response of $|F|_{\mathrm{max}}$.
\begin{figure}
\centering
\begin{tabular}{ccc}
\includegraphics[bb=20bp 20bp 484bp 412bp,clip,width=0.32\textwidth]{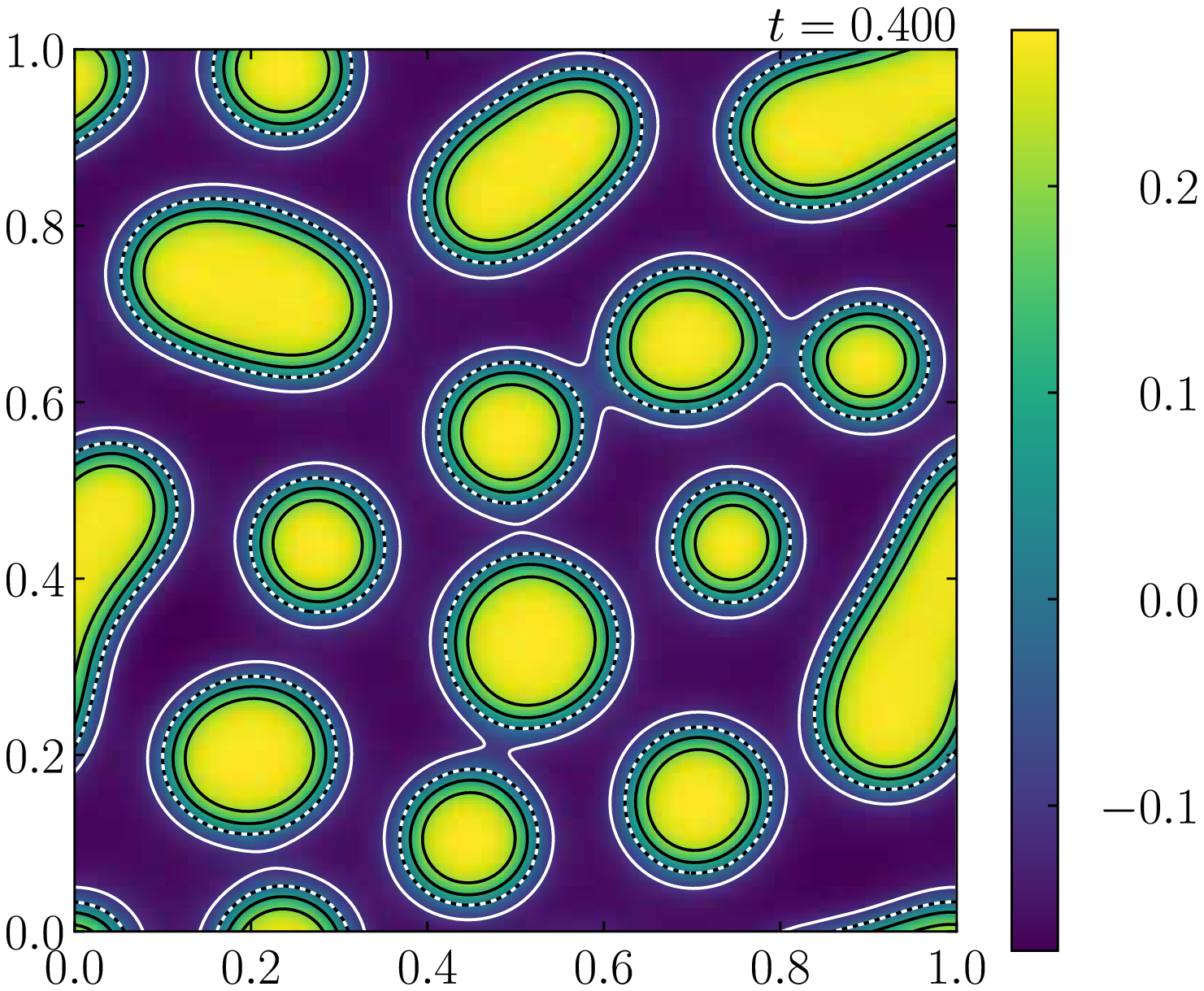}&
\includegraphics[bb=20bp 20bp 484bp 412bp,clip,width=0.32\textwidth]{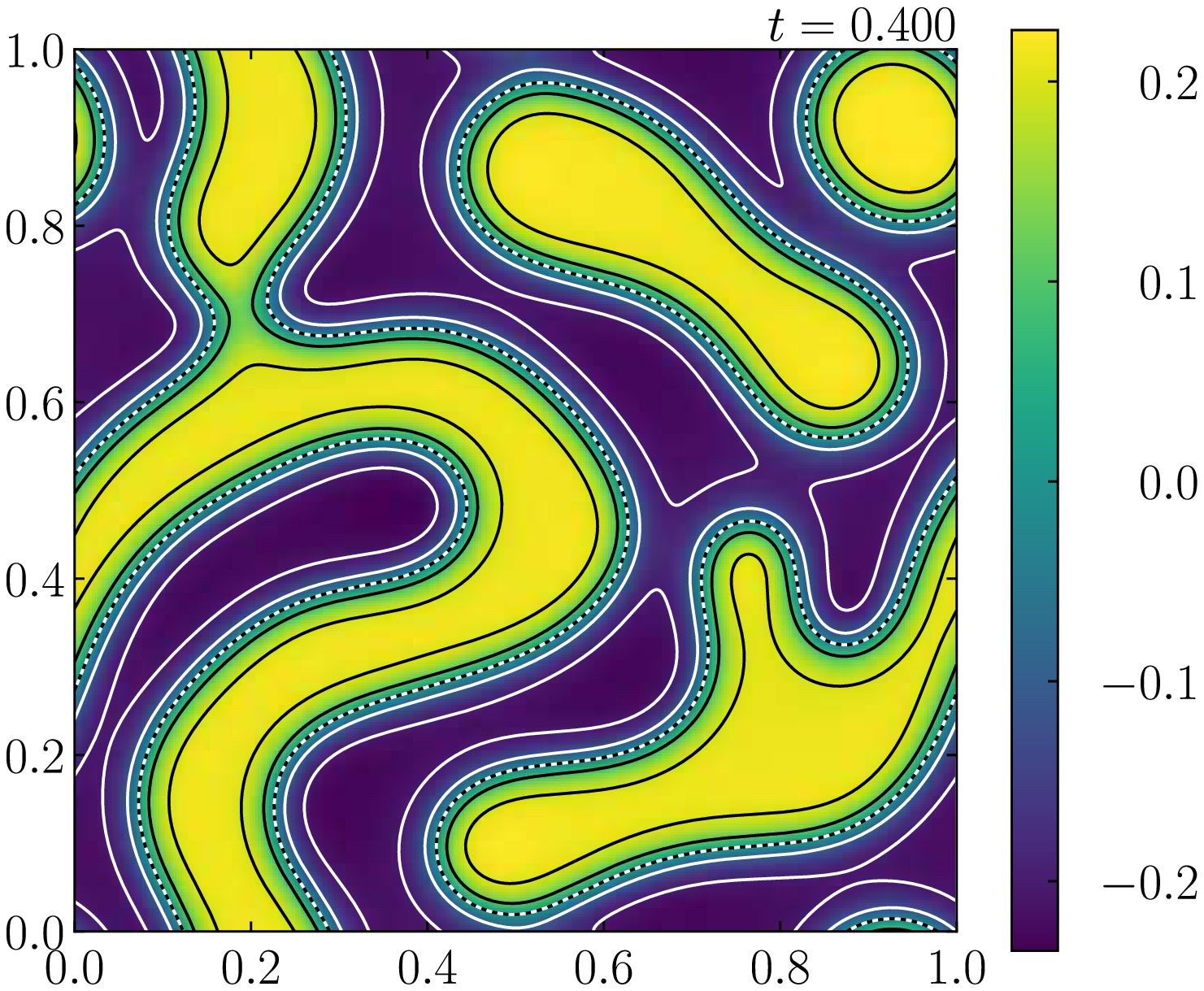}&
\includegraphics[bb=20bp 20bp 484bp 412bp,clip,width=0.32\textwidth]{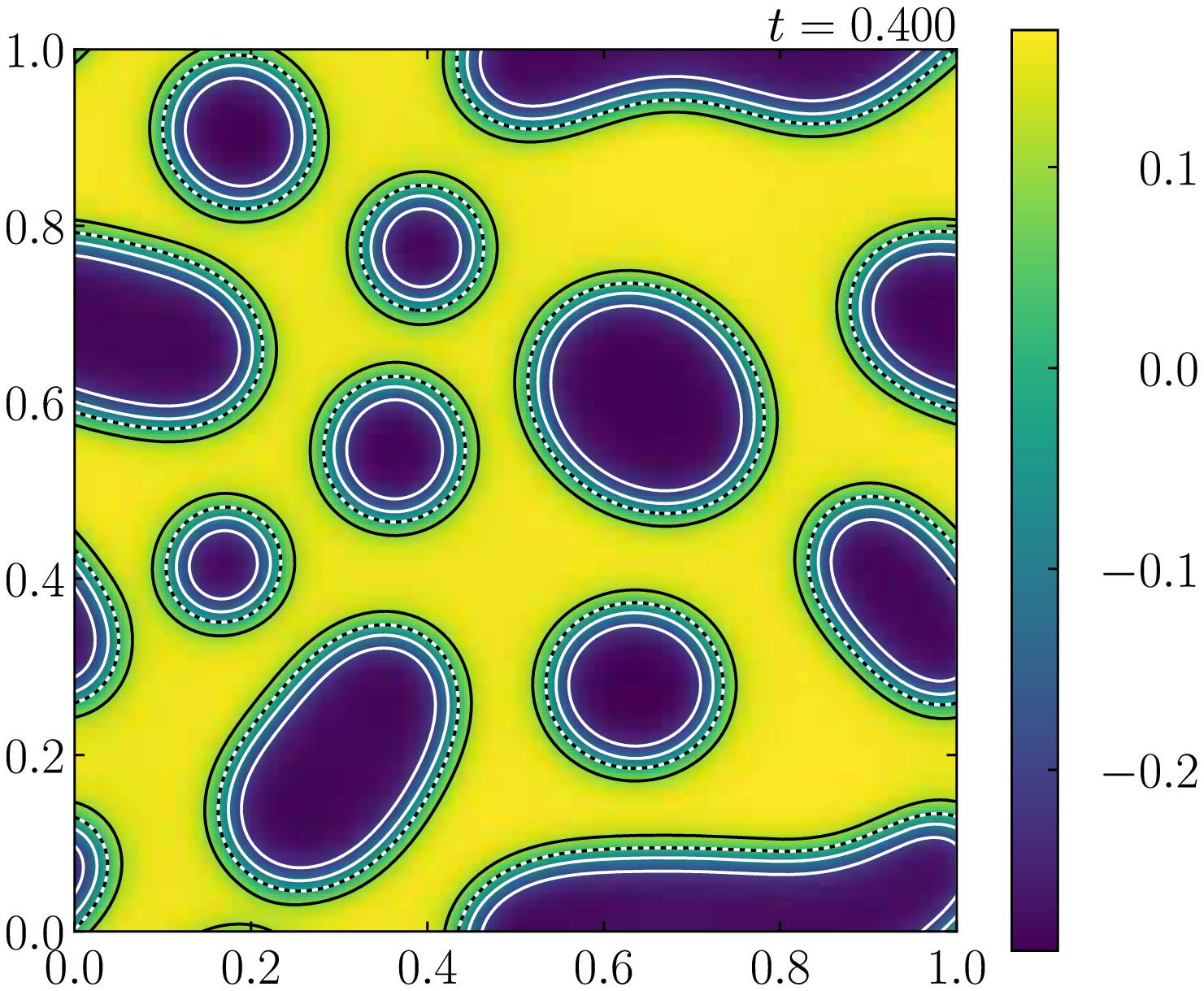}\\
\small (a) $c=3.8$, $\chi_\mathrm{av}=17/60$  & 
\small (b) $c=3.8$, $\chi_\mathrm{av}=1/3$ & 
\small (c) $c=3.8$, $\chi_\mathrm{av}=23/60$ 
\end{tabular}
\caption{Contour plots of the rescaled density $\chi$ on the $xy$-plane:
two coexisting phases at the instance $t=0.400$ induced by an Gaussian
noise (with the standard deviation of 0.001) disturbance of an initial
uniform state. The scale number in the legend indicates the value
of $\chi-\chi_{\mathrm{av}}$. The contours are drawn with the interval
of $0.1$. The contour of $\chi=\chi_{\mathrm{av}}$ is drawn by a
dotted line, while other contours by solid lines. \label{fig:Phase-transition-after}}
\end{figure}
\begin{figure}
\centering
\begin{tabular}{ccc}
\includegraphics[bb=0bp 20bp 422bp 349bp,clip,width=0.4\textwidth]{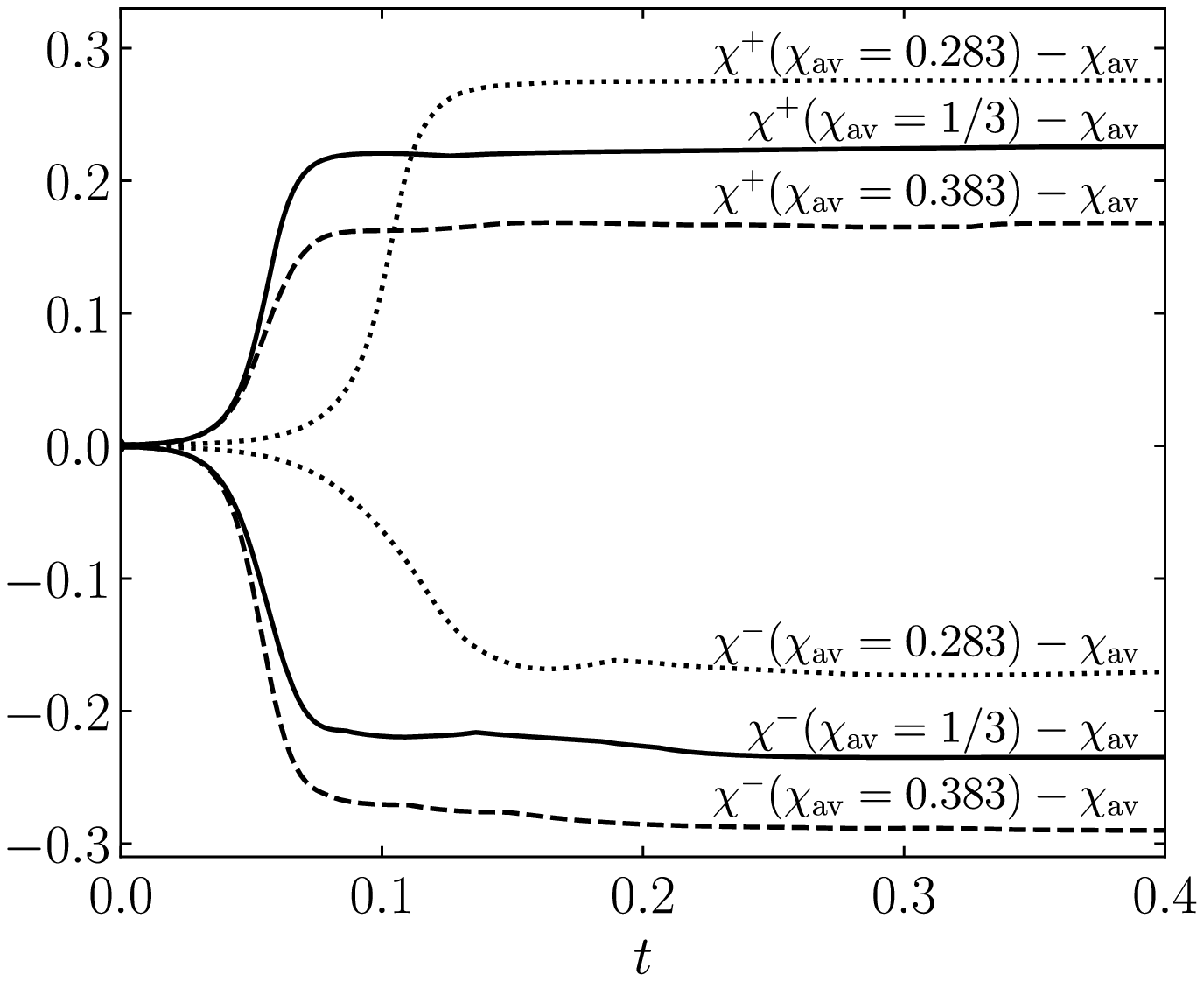}&&
\includegraphics[bb=0bp 10bp 412bp 329bp,clip,width=0.4\textwidth]{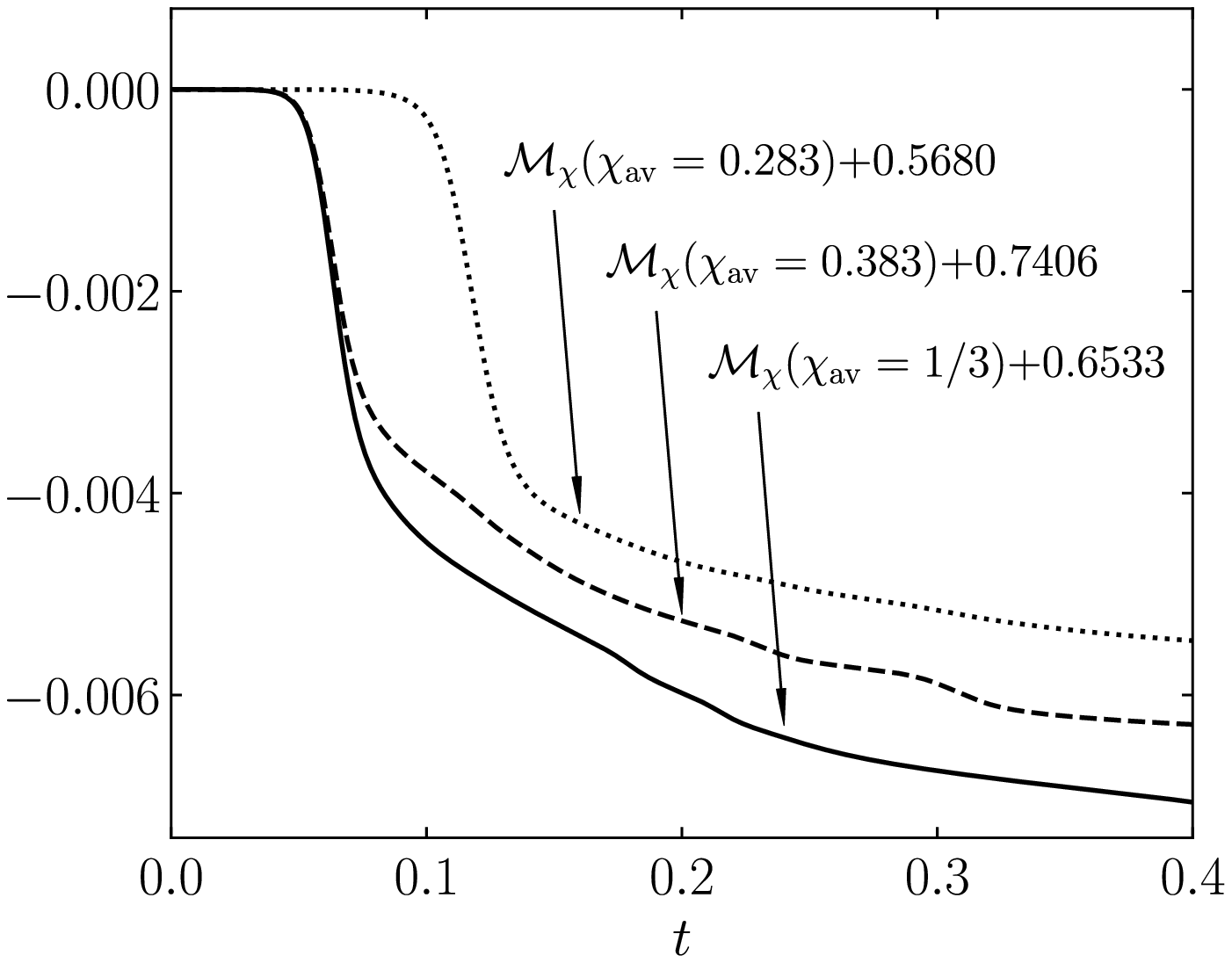}\\
\small (a) && \small (b)\\
\end{tabular}
\caption{Time evolution of the maximum/minimum of $\chi$, say $\chi^{+}$
and $\chi^{-}$, and that of the system total free energy $\mathcal{M}_{\chi}$
in two dimensional cases. (a) $\chi^{\pm}$ vs. $t$ , (b) $\mathcal{M}_{\chi}$
vs. $t$ . Two parameters $K$ and $c$ are commonly set as $K=4.3976\times10^{-5}$
and $c=3.8$, while the values of $\chi_{\mathrm{av}}$ are shown
in the figure. The initial values of $\mathcal{M}_{\chi}$ are $-0.5680$, $-0.6533$,
and $-0.7406$ for $\chi_{\mathrm{av}}=0.283(=17/60),$ $1/3$, and
$0.383(=23/60)$, respectively. \label{fig:Time-evolution-of} }
\end{figure}

In the two dimensional case, we observe a different feature of interface
dynamics, which is absent in the one dimensional case and thus can
be attributed to a multi-dimension effect; see figure~\ref{fig:Phase-transition-after}.
That is, depending on the average $\chi_{\mathrm{av}}$, the formation
of interface geometry changes in quality. When $\chi_{\mathrm{av}}$
is high (low), the regions of dilute (dense) phase appear rather separately;
and occasionally connected dilute (dense) regions change their shape
toward circular discs.
When $\chi_{\mathrm{av}}$ is intermediate, the interface keeps connected
and accordingly its geometry remains complicated. Figure~\ref{fig:Time-evolution-of}
shows the time evolution of the maximum/minimum of $\chi$ and $\mathcal{M}_{\chi}$.
By comparing figures~\ref{fig:Time-evolution-of}(a) and (b), the
main decrease (or first drop) of $\mathcal{M}_{\chi}$ looks triggered
by the first occurrence of phase transition. $\chi^{\pm}$ are almost
saturated during the subsequent gradual decrease of $\mathcal{M}_{\chi}$.
The gradual decrease of $\mathcal{M}_{\chi}$ looks attributed to
a gradual deformation of the interface. 

As to the details of the present numerical computations, the reader
is referred to Appendix~\ref{sec:Numerical-scheme}.

\section{Concluding remark}

In the present paper, we presented a simple kinetic model for the
phase transition of the van der Waals fluid. We constructed the model as
simple as possible with retaining the essential features for reproducing
the phase transition phenomenon. Although our model is rather primitive,
it is reasonable enough to retain a firm connection to fluid dynamical
and statistical mechanical concepts available in the literature. The
simple role of the collision term as a thermal bath makes it easier
to find the monotonically decreasing functional in time  by the H theorem and its relation to
the free energy in thermodynamics. The numerical simulations were
conducted as well for the Cahn\textendash Hilliard type equation that
was obtained in the continuum limit of the presented model. The simulations
demonstrated the actual occurrence of phase transition with this model and
provided some details of dynamics in the near equilibrium regime.

As was briefly mentioned, 
we shall extend the present model to be applicable to far out of equilibrium gas flows.  
In such flows, the isothermal approximation is no longer appropriate 
and the contact with external walls is common.
The extensions in these directions are not straightforward
and are left for future works.

\appendix

\section{Derivations of some equations}\label{sec:DSE}

The equalities (\ref{eq:equivalence}) and (\ref{eq:extforce}) are obtained as follows.
First,  the integration by part results in (\ref{eq:equivalence}):
\begin{equation}
   \int_{D}\rho v_{i}\frac{\partial\Phi_{L}}{\partial X_{i}}d\bm{X}
=  \int_{D}\frac{\partial}{\partial X_{i}}(\rho v_{i}\Phi_{L})d\bm{X}-\int_{D}\frac{\partial \rho v_{i}}{\partial X_{i}}\Phi_{L}d\bm{X}
=  \int_{D}\frac{\partial\rho}{\partial t}\Phi_{L}d\bm{X}, 
\end{equation}
because at the second equality the first term vanishes by the periodic condition and the second term is transformed into the term on the right-hand side by using (7a).
Next, using the definition of  $\Phi_L$ [see (\ref{eq:longrange})], the right-hand side of the above equation is transformed as
\begin{align}
    \int_{D}\frac{\partial\rho}{\partial t}\Phi_{L}d\bm{X} 
=&  \frac{1}{m}\int_{D}d\bm{X}\frac{\partial\rho(\bm{X})}{\partial t}\int_{\mathbb{R}^3} d\bm{r}\,\Psi(|\bm{r}|)\{\rho(\bm{X}+\bm{r})-\rho(\bm{X})\} \nonumber \displaybreak[0]\\
=&  \frac{1}{m}\frac{d}{dt}\int_{D}d\bm{X}\int_{\mathbb{R}^3} d\bm{s}\,\rho(\bm{X})\Psi(|\bm{X}-\bm{s}|)\{\rho(\bm{s})-\rho(\bm{X})\} \nonumber \\
&  -\frac{1}{m}\int_{D}d\bm{X}\int_{\mathbb{R}^3} d\bm{s}\,\rho(\bm{X})\Psi(|\bm{X}-\bm{s}|)\frac{\partial}{\partial t}\{\rho(\bm{s})-\rho(\bm{X})\} \nonumber \displaybreak[0] \\
=&  \frac{1}{m}\frac{d}{dt}\int_{D}d\bm{X}\int_{\mathbb{R}^3} d\bm{s}\,\rho(\bm{X})\Psi(|\bm{X}-\bm{s}|)\{\rho(\bm{s})-\rho(\bm{X})\} \nonumber \\
 & -\frac{1}{m}\int_{D}d\bm{s}\int_{\mathbb{R}^3} d\bm{X}\,\rho(\bm{X})\Psi(|\bm{X}-\bm{s}|)\frac{\partial}{\partial t}\rho(\bm{s}) \nonumber \\
& +\frac{1}{m}\int_{D}d\bm{X}\int_{\mathbb{R}^3} d\bm{s}\,\rho(\bm{X})\Psi(|\bm{X}-\bm{s}|)\frac{\partial}{\partial t}\rho(\bm{X}) \nonumber  \displaybreak[0]\\
=&  \frac{1}{m}\frac{d}{dt}\int_{D}d\bm{X}\int_{\mathbb{R}^3} d\bm{s}\,\rho(\bm{X})\Psi(|\bm{X}-\bm{s}|)\{\rho(\bm{s})-\rho(\bm{X})\} \nonumber \\
&  -\frac{1}{m}\int_{D}d\bm{X}\int_{\mathbb{R}^3} d\bm{s}\,\rho(\bm{s})\Psi(|\bm{X}-\bm{s}|)\frac{\partial}{\partial t}\rho(\bm{X}) \nonumber \\
& +\frac{1}{m}\int_{D}d\bm{X}\int_{\mathbb{R}^3} d\bm{s}\,\rho(\bm{X})\Psi(|\bm{X}-\bm{s}|)\frac{\partial}{\partial t}\rho(\bm{X}) \nonumber  \displaybreak[0]\\
= & \frac{1}{m}\frac{d}{dt}\int_{D}d\bm{X}\int_{\mathbb{R}^3} d\bm{s}\,\rho(\bm{X})\Psi(|\bm{X}-\bm{s}|)\{\rho(\bm{s})-\rho(\bm{X})\} \nonumber \\
 & -\frac{1}{m}\int_{D}d\bm{X}\frac{\partial\rho(\bm{X})}{\partial t}\int_{\mathbb{R}^3} d\bm{s}\,\Psi(|\bm{X}-\bm{s}|)\{\rho(\bm{s})-\rho(\bm{X})\} \nonumber  \displaybreak[0]\\
= & \frac{d}{dt}\int_{D}\rho\Phi_{L}d\bm{X}
   -\int_{D}\frac{\partial\rho}{\partial t}\Phi_{L}d\bm{X}. 
\end{align}
Here, we have suppressed $t$ in the arguments of $\rho$ for brevity.
In the second term just after the third equality, the ranges of integration
with respect to $\bm{X}$ and $\bm{s}$ have been interchanged by using the periodicity in space. 

The above derivation of (\ref{eq:extforce}) relies on the specific form of $\Phi_{L}[\rho]$. 
However, we can show that (\ref{eq:extforce}) is valid as well when $\Phi_{L}=-\kappa\Delta\rho$.
We omit its calculation here. 

The reduction of $\mathcal{M}$ into the form (\ref{eq:defM}) is carried out as follows.
\begin{align}
\mathcal{M}(t) &\equiv \int_{D}\{\langle f\ln\frac{f}{c_{0}}\rangle+\rho\ln(\frac{T^{3/2}}{T_{*}^{3/2}}\frac{\rho_{0}}{\rho})+\frac{\rho}{RT_{*}}(\mathcal{A}+\frac{1}{2}\bm{v}^{2}+\frac{1}{2}\Phi_{L})\}d\bm{X}\nonumber \displaybreak[0] \\
&  =\int_{D}\{\langle f\ln\frac{f}{c_{0}}\rangle+\frac{1}{RT_{*}}(\frac{1}{2}\langle\bm{c}^{2}f\rangle+\frac{1}{2}\rho\bm{v}^{2}+\int\Phi_{S}d\rho)+\frac{\rho}{2RT_{*}}\Phi_{L}[\rho]\}d\bm{X}\nonumber \displaybreak[0] \\
&  =\int_{D}\{\langle f\ln\frac{f}{c_{0}}\rangle+\frac{1}{RT_{*}}(\frac{1}{2}\langle\bm{\xi}^{2}f\rangle+\int\Phi_{S}d\rho)+\frac{\rho}{2RT_{*}}\Phi_{L}[\rho]\}d\bm{X}\nonumber \displaybreak[0] \\
&  =\int_{D}\{\langle f\ln\frac{f}{\rho_{0}M_{*}}\rangle+\frac{1}{RT_{*}}\int\Phi_{S}d\rho+\frac{\rho}{2RT_{*}}\Phi_{L}[\rho]\}d\bm{X}.
\end{align}
In the above transformation, there are two keys: one is the elimination of $\mathcal{A}$ from the expression by using (\ref{eq:Helmholtz}),
and the other is the relation  $\langle \bm{\xi}^2 f\rangle \propto \langle f\ln (\rho_0 M_*/c_0) \rangle$.

The second approximation to $g$, namely (\ref{eq:g_2nd}), is obtained by setting $f=f_{0}+g_{1}$ in (\ref{eq:g}).
The process of transformation is as follows.
\begin{align}
g_{2}=& -\varepsilon(\mathrm{Sh}\frac{\partial f_{0}}{\partial t}+\zeta_{i}\frac{\partial f_{0}}{\partial x_{i}}-\frac{\partial\phi}{\partial x_{i}}\frac{\partial f_{0}}{\partial\zeta_{i}})\frac{1}{A(\rho)}-\varepsilon(\mathrm{Sh}\frac{\partial g_{1}}{\partial t}+\zeta_{i}\frac{\partial g_{1}}{\partial x_{i}}-\frac{\partial\phi}{\partial x_{i}}\frac{\partial g_{1}}{\partial\zeta_{i}})\frac{1}{A(\rho)} \nonumber \displaybreak[0] \\
=& \,g_{1}+\varepsilon^{2}(\zeta_{i}\frac{\partial}{\partial x_{i}}-\frac{\partial\phi}{\partial x_{i}}\frac{\partial}{\partial\zeta_{i}})\{\zeta_{j}\frac{\rho E}{A(\rho)}\frac{\partial}{\partial x_{j}}(\ln\rho+2\phi)\}+o(\varepsilon^{2}) \nonumber \displaybreak[0] \\
=& -\varepsilon\{\mathrm{Sh}\frac{\partial\rho}{\partial t}+\zeta_{i}\rho\frac{\partial}{\partial x_{i}}(\ln\rho+2\phi)\}\frac{E}{A(\rho)}+\varepsilon^{2}\{\frac{\partial}{\partial x_{i}}\{\frac{\rho}{A(\rho)}\frac{\partial}{\partial x_{j}}(\ln\rho+2\phi)\}\}\zeta_{i}\zeta_{j}E\nonumber \\
& +\varepsilon^{2}2\rho\frac{\partial\phi}{\partial x_{i}}\frac{\partial}{\partial x_{j}}(\ln\rho+2\phi)(\zeta_{i}\zeta_{j}-\frac{1}{2}\delta_{ij})E\frac{1}{A(\rho)}+o(\varepsilon^{2}) \nonumber \displaybreak[0] \\
=& -\varepsilon\frac{\rho}{A(\rho)}\frac{\partial}{\partial x_{i}}(\ln\rho+2\phi)\zeta_{i}E+\frac{\varepsilon^{2}}{A(\rho)}\Big\{\frac{\partial}{\partial x_{i}}\{\frac{\rho}{A(\rho)}\frac{\partial}{\partial x_{j}}(\ln\rho+2\phi)\}\nonumber \\
& +\frac{2\rho}{A(\rho)}\frac{\partial\phi}{\partial x_{i}}\frac{\partial}{\partial x_{j}}(\ln\rho+2\phi)\Big\}(\zeta_{i}\zeta_{j}-\frac{1}{2}\delta_{ij})E+o(\varepsilon^{2}),
\end{align}
where (\ref{eq:firstorder}) has been taken into account at the fourth equality.

Finally, the final form of (\ref{eq:H_left}) is the consequence of the following transformation.
\begin{align}
 & \langle(1+\ln\frac{f}{\rho_{0}M_{*}})(\frac{\partial f}{\partial t}+\xi_{i}\frac{\partial f}{\partial X_{i}}+F_{i}\frac{\partial f}{\partial\xi_{i}})\rangle \nonumber \displaybreak[0] \\
= & \frac{\partial}{\partial t}\langle f\ln\frac{f}{c_{0}}\rangle+\frac{\partial}{\partial X_{i}}\langle\xi_{i}f\ln\frac{f}{c_{0}}\rangle+\frac{1}{2RT_{*}}\langle\xi^{2}(\frac{\partial f}{\partial t}+\xi_{i}\frac{\partial f}{\partial X_{i}}+F_{i}\frac{\partial f}{\partial\xi_{i}})\rangle \nonumber \displaybreak[0] \\
= & \frac{\partial}{\partial t}\langle f\ln\frac{f}{c_{0}}\rangle+\frac{\partial}{\partial X_{i}}\langle\xi_{i}f\ln\frac{f}{c_{0}}\rangle+\frac{1}{RT_{*}}\Big\{\frac{\partial}{\partial t}(\frac{1}{2}\langle\bm{c}^{2}f\rangle+\frac{1}{2}\rho v^{2}+\int\Phi_{S}d\rho)+\frac{\partial}{\partial x_i}\{(\frac{1}{2}\langle\bm{c}^{2}f\rangle\nonumber \\
&  +\frac{1}{2}\rho v^{2}+\int\Phi_{S}d\rho)v_{i}+\frac{1}{2}\langle c_{i}\bm{c}^{2}f\rangle+(\langle c_{i}c_{j}f\rangle+\int\rho\Phi_{S}^{\prime}d\rho\delta_{ij})v_{j}\}\Big\}+\frac{\rho v_{i}}{RT_{*}}\frac{\partial\Phi_{L}}{\partial x_i} \nonumber \displaybreak[0] \\
= & \frac{\partial}{\partial t}\Big\{\langle f\ln\frac{f}{c_{0}}\rangle+\frac{\rho}{RT_{*}}\{\frac{1}{2}v^{2}+\mathcal{A}+RT_{*}\ln(\frac{T^{3/2}}{T_{*}^{3/2}}\frac{\rho_{0}}{\rho})\}\Big\}+\frac{\partial}{\partial X_{i}}\Big\{\langle\xi_{i}f\ln\frac{f}{c_{0}}\rangle\nonumber \\
&  +\frac{1}{RT_{*}}\{\rho v_{i}[\frac{1}{2}v^{2}+\mathcal{A}+RT_{*}\ln(\frac{T^{3/2}}{T_{*}^{3/2}}\frac{\rho_{0}}{\rho})]+\frac{1}{2}\langle c_{i}\bm{c}^{2}f\rangle+p_{ij}v_{j}\}\Big\}+\frac{\rho v_{i}}{RT_{*}}\frac{\partial\Phi_{L}}{\partial x_i} \nonumber \displaybreak[0] \\
= & \frac{\partial}{\partial t}\{\langle f\ln\frac{f}{c_{0}}\rangle+\rho\ln(\frac{T^{3/2}}{T_{*}^{3/2}}\frac{\rho_{0}}{\rho})+\frac{\rho}{RT_{*}}(\mathcal{A}+\frac{1}{2}v^{2})\}+\frac{\partial}{\partial X_{i}}\Big\{\langle\xi_{i}f\ln\frac{f}{c_{0}}\rangle\nonumber \\
&  +\rho v_{i}\ln(\frac{T^{3/2}}{T_{*}^{3/2}}\frac{\rho_{0}}{\rho})+\frac{1}{RT_{*}}\{\rho(\mathcal{A}+\frac{1}{2}v^{2})v_{i}+\frac{1}{2}\langle c_{i}\bm{c}^{2}f\rangle+p_{ij}v_{j}\}\Big\}+\frac{\rho v_{i}}{RT_{*}}\frac{\partial\Phi_{L}}{\partial x_i},
\end{align}
where $c_{0}=\rho_{0}(2\pi RT_{*})^{-3/2}$ and (\ref{eq:mass}) has
been taken into account.

\section{Some details of the numerical computations \label{sec:Numerical-scheme}}

The original system is first discretized uniformly in each direction
of space, where the second order central difference is adopted. To
be more precise, the equation (\ref{Cahn}) is discretized in space
as%
\begin{subequations}
\begin{align}
\frac{\partial\chi}{\partial t}(x) & =\delta_{h}[\chi(\tilde{x})\Phi^{\prime}(\chi(\tilde{x}))\delta_{h}[\chi](\tilde{x})-K\chi(\tilde{x})\delta_{h}^{3}[\chi](\tilde{x})](x), \\
 & \delta_{h}[f](x)\equiv\frac{f(x+h)-f(x-h)}{2h}, \\
 & \delta_{h}^{3}[f](x)\equiv\frac{f(x+2h)-2f(x+h)+2f(x-h)-f(x-2h)}{2h^{3}},
\end{align}
\end{subequations}
for one-dimensional (1D) simulations, while%
\begin{subequations}\label{CH2D}
\begin{align}
\frac{\partial\chi}{\partial t}(x,y)= & \Phi^{\prime}(\chi(x,y))\Big\{\{\delta_{hx}[\chi](x,y)\}^{2}+\{\delta_{hy}[\chi](x,y)\}^{2}+\chi(x,y)\delta_{h}^{2}[\chi](x,y)\Big\}\nonumber \\
 & +\Phi^{\prime\prime}(\chi(x,y))\chi(x,y)\Big\{\{\delta_{hx}[\chi](x,y)\}^{2}+\{\delta_{hy}[\chi](x,y)\}^{2}\Big\}\nonumber \\
 & -K\delta_{h}^{2}[\delta_{h}^{2}[\chi](\tilde{x},\tilde{y})](x,y), \\
 & \delta_{hx}[f](x,y)\equiv\frac{f(x+h,y)-f(x-h,y)}{2h}, \\
 & \delta_{hy}[f](x,y)\equiv\frac{f(x,y+h)-f(x,y-h)}{2h}, \\
 & \delta_{h}^{2}[f](x,y)\equiv\frac{f(x+h,y)-2f(x,y)+f(x-h,y)}{h^{2}}\nonumber \\
 & \qquad\qquad\qquad+\frac{f(x,y+h)-2f(x,y)+f(x,y-h)}{h^{2}},
\end{align}
\end{subequations}%
for two-dimensional (2D) simulations. Here $h$
is the interval of the uniform grid and $t$ has been suppressed in
the argument of functions. In the standard grid system, the spatial
domain is divided into 800 uniform intervals in each direction. All
the results shown in section~\ref{subsec:Numerical-simulation-of} are
those obtained by the computations with the standard grid. As is already
mentioned in section~\ref{subsec:Numerical-simulation-of}, the initial
disturbance for each simulation is commonly a Gaussian noise with
the standard deviation of $0.001$, but it is shifted in amplitude
so as not to change the total mass in the domain. Furthermore, the
Gaussian noise was generated on the basis of 100 grid for 1D and $100\times100$
grid for 2D simulations so as not to change the initial disturbance
for different grid systems. The minimum length of the generated randomness
is eight-times longer than the interval of the standard grid ($800$
for 1D and $800\times800$ for 2 D) in each spatial direction. This
rather artificial care enables us to check the grid convergence of
the numerical solutions, with keeping the randomness of the initial
disturbance.

The time integration of the discretized system for 1D has been carried
out by implementing the double-precision version of LSODA code in
the ODEPACK developed by the Lawrence Livermore National Laboratory,
which is available from http://www.netlib.org/odepack/ as of August
22, 2017. The code uses the Adams (predictor-corrector) method in
the nonstiff case and the Backward Differentiation Formula (BDF) method
in the stiff case, and it is decided adaptively which method to use.
Actually, however, the Adams method was used only at the first time
step in all of our simulations. The code uses both the variable timestep
and the multistep, and the size of timestep and the degree of multistep
(up to four steps) are optimized automatically as well. For the details
of related optimization principle and features of LSODA itself, the
reader is referred to \cite{HNW87}, as well as the summary
text ``odkd-sum'' in the ODEPACK.

In the meantime, the time integration of the discretized system for
2D has been carried out by the explicit two-steps Runge\textendash Kutta
method, which is of the second order accuracy. If we symbolically
rewrite (\ref{CH2D}) as $\partial\chi/\partial t=F(\chi)$, the time
integration has been carried out by the following set of the prediction
and correction steps%
\begin{subequations}
\begin{align}
 & \tilde{\chi}_{n+1}=\chi_{n}+F(\chi_{n})\Delta t_{n+1},\\
 & \chi_{n+1}=\chi_{n}+\frac{1}{2}\{F(\chi_{n})+F(\tilde{\chi}_{n+1})\}\Delta t_{n+1},
\end{align}
\end{subequations}%
where $\chi_{n}$ denotes the value of $\chi$
at $t=t_{n}$ ($t_{0}=0$) and $\Delta t_{n+1}=t_{n+1}-t_{n}$. The
correction step is taken only once in a single time step, namely the
so-called PEC mode is adopted. The timestep $\Delta t_{n+1}$ is fixed,
in contrast to 1D simulations, as $1\times10^{-9}$ for the standard
grid ($800\times800$), $2\times10^{-8}$ for $400\times400$ grid,
$2\times10^{-7}$ for $200\times200$ grid, and $2\times10^{-6}$
for $100\times100$ grid. 
\begin{figure}
\centering
\begin{tabular}{cc}
\includegraphics[bb=10bp 15bp 360bp 275bp,clip,width=0.45\textwidth]{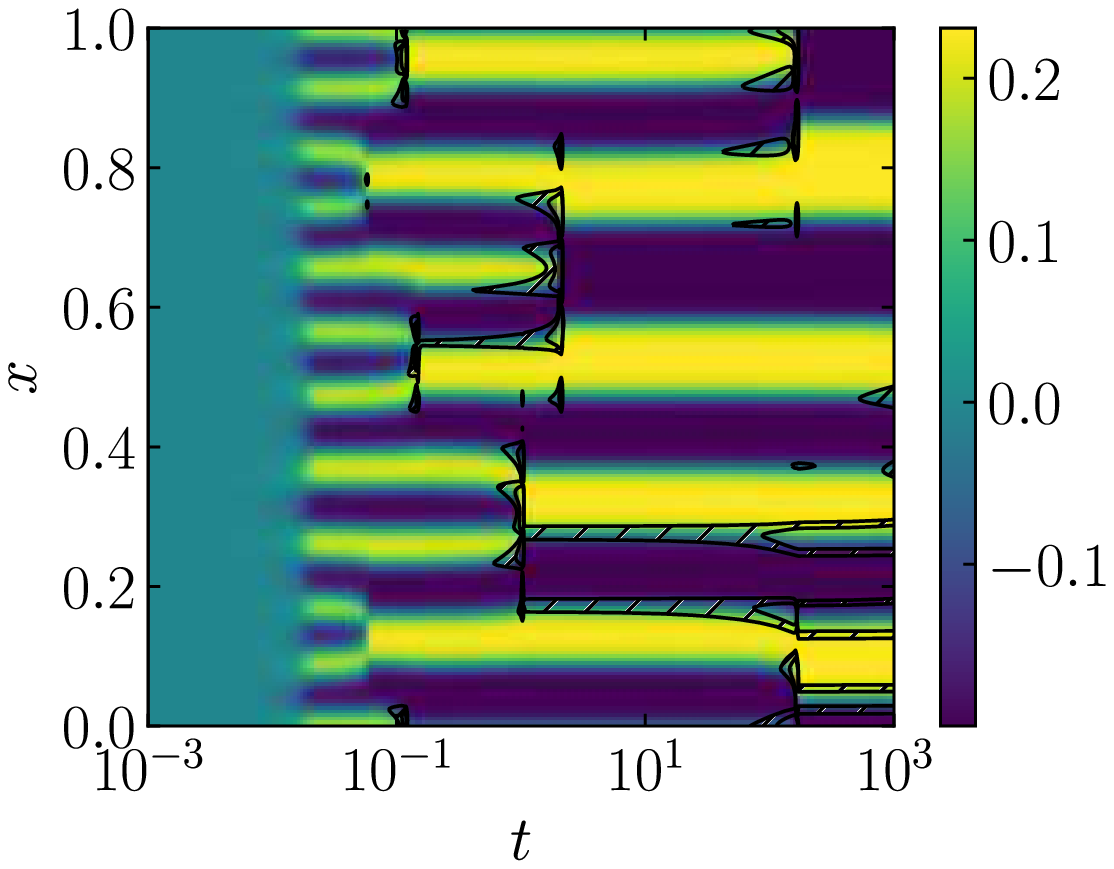}&
\includegraphics[bb=10bp -10bp 442bp 345bp,clip,width=0.4\textwidth]{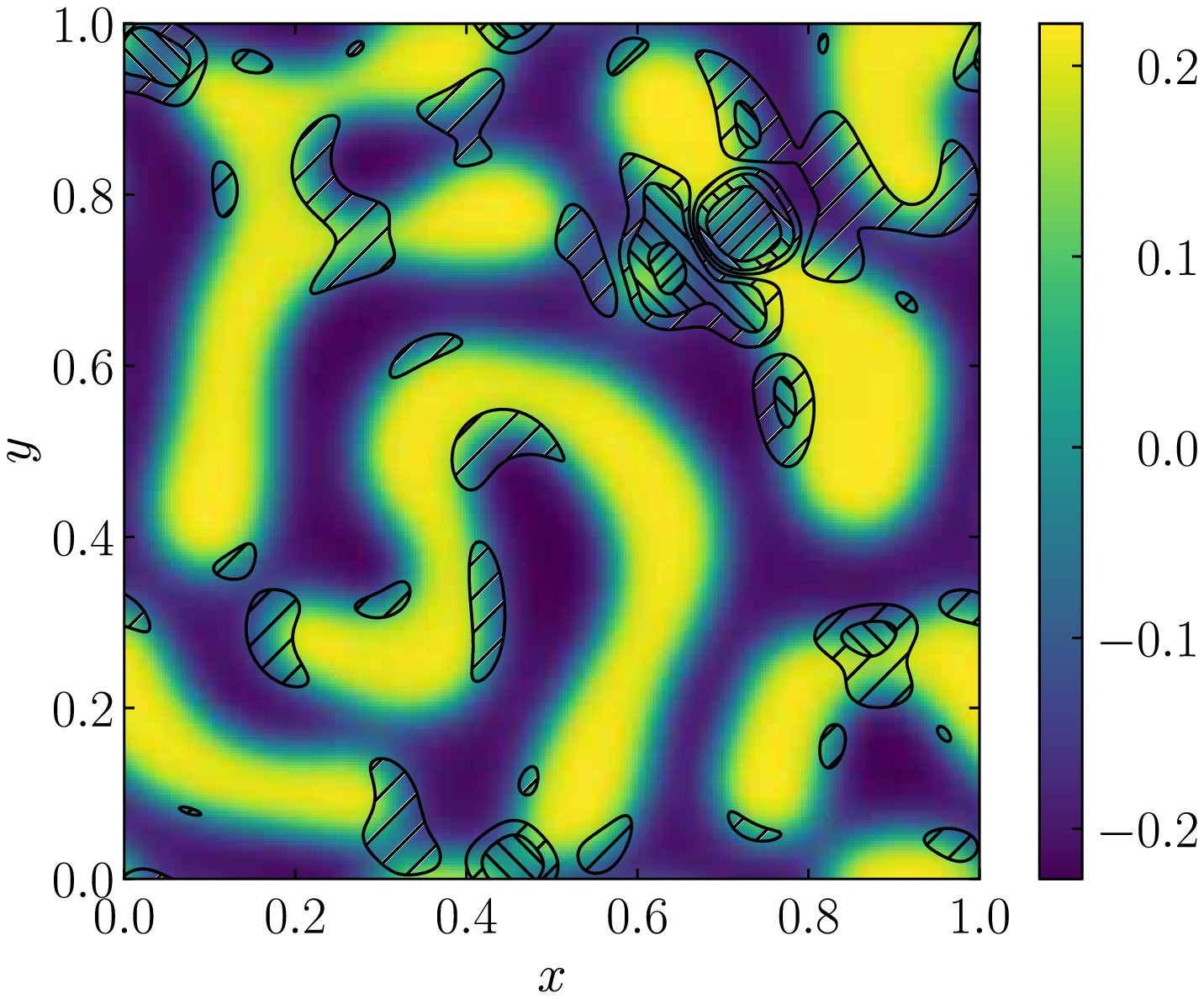}\\
\small (a) 1D & \small (b) 2D \\ 
\end{tabular}
\caption{Grid dependence of the results for the case $c=3.8$, $\chi_{\mathrm{av}}=1/3$,
and $K=4.3976\times10^{-5}$. (a) The difference $\Delta$ between
the results with standard and double-size coarse grids in the 1D simulation.
(b) The difference $\Delta$ at $t=0.2$ between the results with
standard and double-size coarse grids in the 2D simulation. The three
types of hatched area, namely the upward hatched area with wide interval,
the downward hatched area, and the upward hatched area with narrow
interval, represent the area where $0.003<\Delta<0.01$, $0.01<\Delta<0.03$,
and $0.03<\Delta$. The scale in the legend shows $\chi-\chi_{\mathrm{av}}$.
\label{fig:Grid-dependence-of}}
\end{figure}
 We implemented the LSODA code as well for the time integration in
2D. However, it turned out to be very time consuming and had to be
limited only to four- or more-times coarse grids. For the four- or
more-times coarse grids, we had a reasonable agreement between the
results of Runge-Kutta and LSODA codes.

The present scheme for both 1D and 2D is not based on a mass preserving
method. Nevertheless, we observed that the total mass was perfectly
preserved in 1D simulations. In contrast, in 2D simulations, a straightforward
implementation caused a gradual change of the total mass in the domain
in both the Runge-Kutta and LSODA codes, which could affect the main
feature of the phase transition in the system. We therefore renormalize
the total mass at the beginning of every time step. The adverse side
effect of this remedy should be carefully assessed. We thus performed
the simulations without renormalization for the same grid and those
with renormalization for a more refined grid as well. The multiplied
factor for the renormalization was close to unity, the deviation from
which was about $1.47\times10^{-11}$ for the standard grid ($800\times800$),
$5.50\times10^{-10}$ for $400\times400$ grid, and $2.15\times10^{-8}$
for $200\times200$ grid. The size of deviations per unit time was
decreasing from $0.10$ (for $200\times200$ grid) to $0.028$ or
$0.015$ (for $400\times400$ or $800\times800$ grid), showing the
improvement of reliability by a grid refinement. We did not find any
qualitative difference among the above three types of simulations,
such as the spatial arrangement of different phases, the number of
the dilute/dense regions. However, due to slight differences of the
instance of merging and of the interface position, the grid dependence
of $\chi$ at a fixed position and time is not necessarily small,
see figure~\ref{fig:Grid-dependence-of}. All the numerical results
presented in section~\ref{subsec:Numerical-simulation-of} are those
obtained with the standard grid and with the remedy of total mass
renormalization.

\acknowledgement
The present work was supported in part by JSPS KAKENHI Grant Number
17K18840 and by JSPS and MAEDI under the Japan-France Integrated Action Program (SAKURA).

\end{document}